\documentclass[a4paper,11pt]{article}                                                       
\usepackage{listings}
\usepackage{multicol}
\usepackage{color}

\usepackage{geometry}                                                                                 
\usepackage{amsmath}
\usepackage{amsthm}
\usepackage{dsfont}
\usepackage[utf8]{inputenc}   
\usepackage{graphicx}
\usepackage[blocks]{authblk}

\theoremstyle{plain}
\newtheorem{thm}{Theorem}[section]
\newtheorem{lem}[thm]{Lemma}
\newtheorem{prop}[thm]{Proposition}
\newtheorem*{cor}{Corollary}

\theoremstyle{definition}
\newtheorem{defn}{Definition}[section]
\newtheorem{conj}{Conjecture}[section]

\theoremstyle{remark}
\newtheorem*{rem}{Remark}

\title{Algorithmic Aspects of Switch Cographs}
\author[*]{Vincent Cohen-Addad}
\author[*]{Michel Habib}
\author[*]{Fabien de Montgolfier}
\affil[*]{LIAFA - Université Paris Diderot - Paris 7, 75205 Paris Cedex 13, France.}

\date{}

\begin{document}

\maketitle
\abstract{This paper introduces the notion of involution module, the first generalization of the modular 
  decomposition of 2-structure which has a unique linear-sized decomposition tree.
  We derive an $\mathcal{O}(n^2)$ decomposition algorithm and we take advantage of the involution modular
  decomposition tree to state several algorithmic results.
  Cographs are the graphs that are totally decomposable w.r.t modular decomposition. In a similar way, we introduce the class of switch cographs, 
  the class of graphs that are totally decomposable w.r.t involution modular decomposition.
  This class generalizes the class of cographs and is exactly the class of (Bull, Gem, Co-Gem, $C_5$)-free graphs.
  We use our new decomposition tool to design three practical algorithms for the maximum cut, vertex cover and vertex separator problems.
  The complexity of these problems was still unknown for this class of graphs.
  This paper also improves the complexity of the maximum clique, the maximum independant set, the
  chromatic number and the maximum clique cover problems by giving efficient algorithms, thanks to the decomposition tree.
  Eventually, we show that this class of graphs has Clique-Width at most 4 and
  that a Clique-Width expression can be computed in linear time.
}

\section*{Introduction}
Modular decomposition has arisen in different contexts as a very natural operation on many discrete structures such as
graphs, directed graphs, 2-structures, automata, boolean functions, hypergraphs, or matroids. In graph theory, the study of
modular decomposition as a graph decomposition technique was first introduced by Gallaï \cite{gallai}.
This notion has led to state several important properties of both structural and algorithmic flavour. Many graph classes
such as cographs, $P_4$-sparse graphs or $P_4$-tidy graphs are characterized by the properties of their modular decomposition (see for 
example \cite{graphclasses_survey}). 

Also, several classical graph problems (NP-complete in the general case) can be solved in polynomial time when restricted to classes of 
graphs that are ``\emph{decomposable enough}''. For example, \cite{corneil_P4,cograph_pathcover} designed efficient algorithms for the class of cographs  
which rely on the modular decomposition tree of the cographs. 

We start from a generalization of modular decomposition, namely the umodular decomposition defined in \cite{umodules}.
In his PhD thesis\cite{binhminh_phd}, Bui Xuan has shown that the family of umodules of more general combinatorial objects (such as 2-structures \cite{2-struct})
has no polynomial-sized tree representation. Therefore, as far as we know, there is no generalization of modular decomposition that have
a polynomial-sized tree representation in a more general context than graphs.

In this paper, we introduce the notion of \emph{involution modules}, which is a generalization of modules but a restriction of umodules, and we show that the family of involution modules
of any 2-structure has very strong properties. These properties are similar to the properties of modules, and lead us to derive in $\mathcal{O}(n^2)$ time a unique linear-sized 
decomposition tree for any 2-structure. To this aim we use a very interesting switch operator that generalizes to 2-structures
the well-known Seidel Switch introduced by \cite{Seidel_Switch} and widely studied by \cite{hertz,kratochvil,hayward}.

Then we focus our study on the particular case of 2-structure with two colors, namely undirected graphs which are more concerned 
by the algorithmic aspects than 2-structures.
We consider the class of graphs totally decomposable with respect to the involution modular decomposition.
We call this class the class of \emph{Switch Cographs} and we show that switch cographs are exactly the graphs 
with no induced Gem, Co-Gem, $C_5$ nor Bull subgraphs. This graphs family is already known in the litterature 
(see for example \cite{hertz}) and generalizes the widely studied class of cographs. 
Like the modular decomposition for cographs, the involution modular decomposition provides crucial algorithmic properties for the class of switch cographs.
Using our decomposition approach we give efficient and practical algorithms for the class of switch cographs 
to  well-known graph problems (NP-complete in the general case), namely 
the maximum cut and the vertex separator problems. The complexity of these problems was still unknown for this class of graphs.
Since the Clique-Width of the switch cographs is bounded, the complexity of several graph problems depended on the celebrated Courcelle's theorem.
The theorem implies in particular that the maximum clique, the maximum independant set, the chromatic number the vertex cover and the minimum clique cover problems
can be solved in polynomial time for the class of Switch Cographs.
Nevertheless, the theorem induces a huge constant factor in the big-O notation and cannot be considered of practical interest.
We then show that the involution modular decomposition tree can be used in order to derive a 
Clique-Width expression in linear time leading to a linear-time complexity for these problems. 
Then, we give easily implementable algorithms which ensure the same optimal complexity.
Finally, we conclude this paper by showing that this class of graphs is strictly included in the class of graphs with Clique-width at most 4.

The paper is organized as follows, section \ref{def} recalls definitions and the general framework of modular decomposition, section \ref{involution_modules} introduces
the notion of involution modules, studies its properties and presents the decomposition algorithm. Section \ref{switch_cographs} is devoted to the study of switch cographs
and to the algorithms we designed thanks to the involution modular decomposition.
Eventually, we discuss the noteworthy outcomes and open questions that follow from our work.

\section{Definitions}
\label{def}
We recall some definitions about generalisations of modular decomposition (as they are given in \cite{umodules}).
Let $X$ be a finite set. 
We say that two subsets $A,B \subseteq X$ are \emph{overlapping} if the sets $A \cap B$, $A \setminus B$, $B \setminus A$ are not empty.
Finally, we say that two sets $A,B \subseteq X$ are \emph{crossing} if they are overlapping and $X \neq A \cup B$.é

\begin{defn}{\cite{2-struct} \textbf{2-structure.}
    A 2-structure $G$ is a couple $(X,E)$ where
    $X$ is a finite set (the set of the \emph{vertices}) and $E$ is a function,
    $E : X^2 \rightarrow \mathds{N}$.
  }
\end{defn}

We say that a 2-structure $G$ is symmetric if for all $x,y \in X$, $E(x,y) = E(y,x)$.
An edge over $X$ is a pair $(x,y)$, $x,y \in X$ and $x \neq y$ and let $E_2(X)$ denotes the set of all edges
over $X$. Throughout this paper, we only consider symmetric 2-structures and we always omit the word ``symmetric''.
For a given 2-structure $G=(X,E)$ we say that the set $C = \{i $ $|$ $\exists u,v$ s.t $E(u,v) = i\}$ is the set
of the colors of the 2-structure.
By $N^{i}_{s}(X')$ we denote the set $\{x$ $|$ $x \in X'$ and $E(s,x) = i \}$, basically the set of elements in $X'$ that 
are connected to $s$ with the color $i$.

The reader may remark that any undirected graph is basically a 2-structure with 2 colors.
Let us recall below the usual notation of modular decomposition.

\subsection{Homogeneous Relation, Modules and Umodules}


We now recall the notion of module for a 2-structure.

\begin{defn}{\cite{ehrenfeucht} \textbf{Modules}.
    Let $G=(X,E)$ be a 2-structure.
    A subset $M \subseteq X$ is a module of $G$ if :
\begin{center}
  $\forall m$, $m' \in M$, $\forall i \in C$, $N_m^i(X \setminus M) = N_{m'}^i(X \setminus M)$.
\end{center}
}

\end{defn}

We say that a module $M$ is trivial if $|M|$ $\leq$ $1$ or $M = X$.
We now present the primary properties of modular decomposition. Throughout this section, we denote by
$2^X$ the family of subsets of any finite set $X$.

\begin{defn}{\textbf{Partitive family.}
    Let $X$ be a set of elements.
    $\mathcal{F} \subseteq 2^{X}$ is a partitive family if $X$ and $\emptyset$ $\in \mathcal{F}$ and for any overlapping sets $A$, $B \in \mathcal{F}$, 
    $A \cap B \neq \emptyset$ and $A \cup B \neq X$ implies $A \cap B \in \mathcal{F}$, $A \cup B \in \mathcal{F}$, $A \setminus B \in \mathcal{F}$
    and $A \bigtriangleup B \in \mathcal{F}$.} 
\end{defn}

\cite{habib_decomp} showed that the family of modules of any graph (i.e 2-structure with two colors) is a partitive family
and demonstrated the following theorem of particular importance.

\begin{thm}{\cite{habib_decomp} \textbf{Decomposition theorem of partitive families.}}
    If $F$ is a partitive family, there exists a unique rooted undirected 
    tree-representation of $F$, $\mathcal{T}(F)$, of size $\mathcal{O}(|X|)$.
    This tree representation is such that the internal nodes of $\mathcal{T}(F)$ can be labelled \emph{complete} or \emph{prime}
    such that:
    \begin{itemize}
      \item The leaves are exactly the elements of $X$;
      \item Let $N$ be a node with $k$ siblings $N_1,...,N_k$,
        \begin{description}
        \item If $N$ is a complete node, for any $I \subset \{1, . . . , k\}$ such that $1 < |I| < k$, $\bigcup\limits_{i \in I} X_i \in F$, and
        \item if $N$ is a prime node, for any element $i \in \{1, . . . , k\}$, $ X_i \in F$,\\
          where $X_i$ is the set of elements of leaves whose paths to $N$ traverse $N_i$;
        \end{description}
    \item There are no more sets in $F$ than the ones described above.
    \end{itemize}

\end{thm}

\cite{ehrenfeucht} presented an $\mathcal{O}(|X|^2)$ algorithm which computes the tree-decomposition of the family
of modules of any 2-structure.
We conclude this section by reminding a generalization of modular decomposition introduced by \cite{umodules}. 

\begin{defn}{\textbf{Umodules}.
  Let $G=(X,E)$ be a 2-structure.
  A subset $U$ of $X$ is a umodule if $\forall u,u' \in U,$ $\forall x,x' \in X \backslash U$,\\
  \begin{center}
    $\exists i \in C$, $x \in N_u^i$ and $ x' \notin N_u^i$ $\iff$ $\forall j \in C$, $x \in N_u^j$ and $ x' \notin N_u^j$
  \end{center}}
\end{defn}

This led \cite{umodules} to introduce the notion of partitive crossing family, namely :

\begin{defn}{
    \textbf{Partitive crossing family.}
    Let $X$ be a set of elements.
    $\mathcal{F} \subseteq 2^{X}$ is a partitive crossing family if $X$ and $\emptyset$ $\in \mathcal{F}$ and
    for any crossing sets $A$, $B \in \mathcal{F}$, 
    $A \cap B \neq \emptyset$ and $A \cup B \neq X$ implies $A \cap B \in \mathcal{F}$, $A \cup B \in \mathcal{F}$, $A \setminus B \in \mathcal{F}$
    and $A \bigtriangleup B \in \mathcal{F}$.}
\end{defn}

Then \cite{umodules} showed that the family of umodules of a graph is a partitive crossing family.

\begin{thm}{\cite{BHR} \textbf{Decomposition theorem partitive crossing families.} 
    If $\mathcal{F}$ is a partitive crossing family, there exists a unique unrooted and directed 
    tree-representation of $\mathcal{F}$, $\mathcal{T}(\mathcal{F})$, of size $\mathcal{O}(|X|)$.
    This tree-representation is such that the nodes of $\mathcal{T}(\mathcal{F})$ can be labelled \emph{complete} or \emph{prime}.
    such that:
    \begin{itemize}
    \item For any nodes $N_1,N_2$, if $(N_1,N_2)$ is an arc of the tree then $N_1$ is in the family. 
    \item If $N$ is a node with $k$ in-neighbors $N_1,...,N_k$ :
      \begin{description}
      \item If $N$ is a complete node, for any $I \subset \{1, . . . , k\}$ such that $1 < |I| < k$, $\bigcup\limits_{i \in I} X_i \in F$, and
      \item if $N$ is a prime node, for any element $i \in \{1, . . . , k\}$, $ X_i \in F$,\\
        where $X_i$ is the set of leaves whose paths to $N$ traverse $N_i$.
      \end{description}
    \item There are no more sets in $F$ than the ones described above.
    \end{itemize}
  }
  \label{decomp_theorem}
\end{thm}



\cite{umodules} presented an algorithm which computes for any graph $G=(X,E)$ the tree representation of its
family of umodules with an $\mathcal{O}(|X| + |E|)$ complexity.

\begin{figure}
  \begin{center}
    
    \includegraphics[scale=0.6]{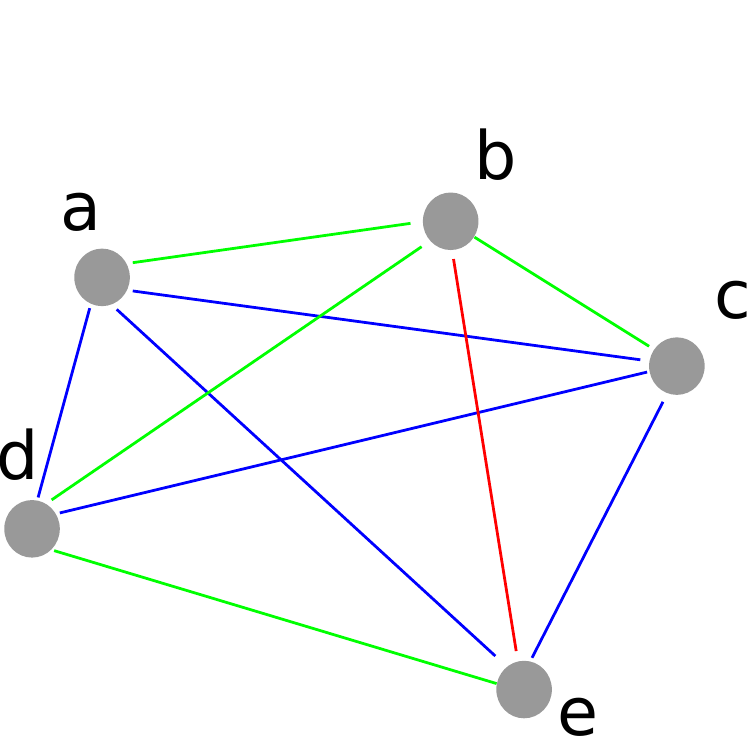}
    \caption{The sets $\{a,c,e\}$ and $\{a,d,e\}$ are crossing umodules
      but their intersection, $\{a,e\}$ is not a umodule.}
    \label{anti_inter}
    
  \end{center}
\end{figure}

\section{Involution Modules, a New Decomposition Tool}
\label{involution_modules}
\subsection{Discussion}
The notion of umodule presented above and due to \cite{umodules,bijoin} induces 
a family which has strong properties of both algorithmic and structural flavour on graphs. 
Nevertheless, unlike the modules, the family of umodules of a 2-structure has no 
polynomial-sized tree-representation and so
cannot be used in order to decompose more general objects such as 2-structures \cite{binhminh_phd}.

For example, figure \ref{anti_inter} shows that there exists a 2-structure with only 3 colors 
whose family of umodules is not closed under intersection.
We found two other 2-structures with 3 colors whose families of umodules are not closed under
difference and symmetric difference.
Eventually, \cite{binhminh_phd} showed that the family of umodules of any 2-structure can not
be represented in polynomial time.
We introduce below the notion of \emph{involution module}, a generalization of modules and a restriction of umodules.
We show that the family of involution
modules of any 2-structure has similar properties as the family of modules, namely the closure
under union, intersection, difference and symmetric difference of crossing sets.
These properties lead to a unique linear-sized tree-representation by theorem \ref{decomp_theorem} and allow us to derive an optimal algorithm that computes it.

\subsection{Definition and Properties}
\begin{defn}{\textbf{Involution Modules.}}
  Let $G=(X,E)$ be a 2-structure, $\mathcal{I}$ an involution of the colors without fix point.
  $U \subset X$ is an involution module if, for all $u,v \in U$, 
  \begin{center}
    \begin{itemize}
    \item Either, $\forall i \in \{1,...,|C|\}$, $N_s^i(X \backslash U) = N_u^{\mathcal{I}(i)}(X \backslash U)$.
    \item Or, $\forall i \in \{1,...,|C|\}$, $N_s^i(X \backslash U) = N_u^{i}(X \backslash U)$.
    \end{itemize}
  \end{center}
\end{defn}

Remark like elements of a umodule, elements of the involution module have to partition the rest of the 2-structure
in the same way.

\begin{figure}
\begin{center}

\includegraphics[scale=0.6]{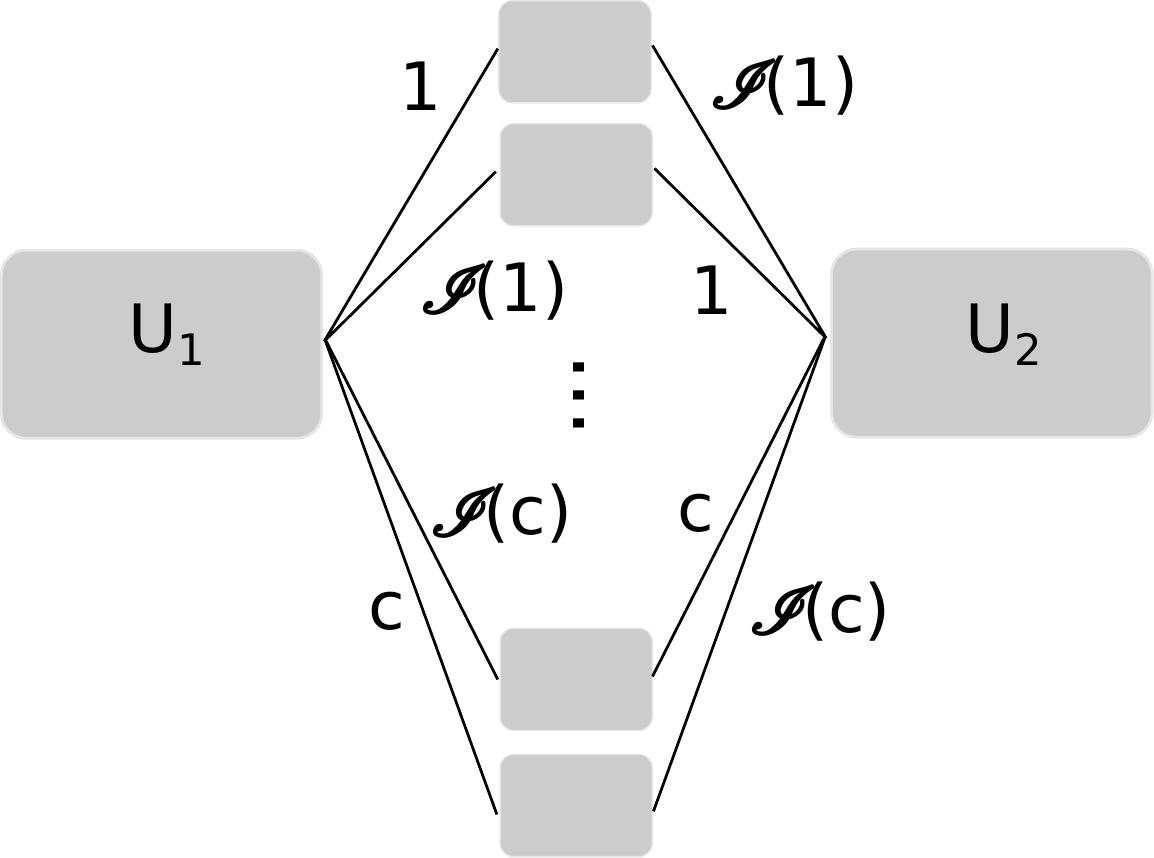}
\caption{An involution module (with respect to an involution $\mathcal{I}$) 
  can be divided into two parts, here parts $U_1$ and $U_2$, such that for each element $e \in U_i$,
  for any neighbor $n$ of $e$ in the rest of the 2-structure, every other element of $U_i$ is connected
  to $n$ with the color $E(e,n)$ and every element of $U_{3-i}$ is connected to $n$ with the color $\mathcal{I}(E(e,n))$,
  i.e the image of $E(e,n)$ by the involution $\mathcal{I}$.}
\label{vision}
\end{center}
\end{figure}

Throughout this paper we will consider involutions without fix point.
Figure \ref{vision} shows how an involution module is connected to the rest of the 2-structure.

Let us now highlight important properties of involution modules.
We begin with a strong characterization property which will be used 
in order to prove the tree-decomposition theorem.

\begin{figure}
  \begin{multicols}{2}
    \begin{center}
      \includegraphics[scale=0.6]{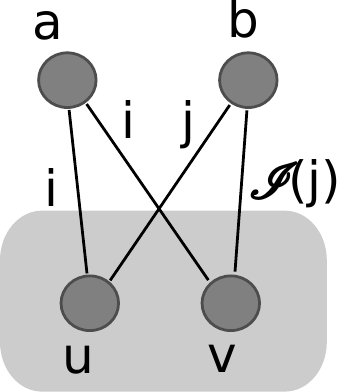}
    \end{center}
    \caption{
      First forbidden pattern for an involution module. There is no involution module which contains $u$ and $v$ and neither $a$ nor $b$.}
    \label{forbidden1}
    
    \begin{center}
      \includegraphics[scale=0.6]{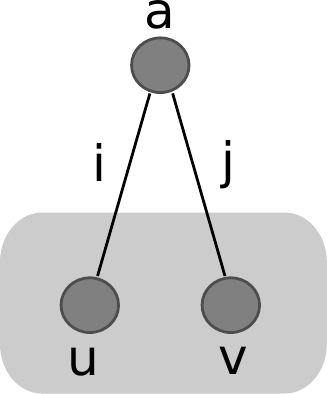}
    \end{center}
    \caption{
      Second forbidden pattern for an involution module. There is no involution module which contains $u$ and $v$ and not $a$ 
      (here $j \neq \mathcal{I}(i)$).}
    \label{forbidden2}
  \end{multicols}
\end{figure}

\begin{prop}{\textbf{Characterization by forbidden patterns}.
    Let $G=(X,E)$ be a 2-structure with $C$ colors, $\mathcal{I}$ an involution of the colors
    and $U \subset X$.\\
    \begin{center}      
      $U$ is an involution module of $G$ $\iff$ $\forall u,v \in U$, $\forall a,b \in X \backslash U$, $\forall i,j \in \mathcal{C}$, 
      figures \ref{forbidden1} and \ref{forbidden2} are not induced in $G$.
    \end{center}
  }
  \begin{proof}
    The \emph{only if} part is easy: $U$ being an involution module, figure \ref{forbidden1} contradicts the two conditions and
    figure \ref{forbidden2} does not abide by the involution.

    Let us now show the \emph{if} part.
    Assume towards contradiction that $\forall u,v \in U$, $\forall a,b \in X \backslash U$, $\forall i,j \in \mathcal{C}$, 
    figures \ref{forbidden1} and \ref{forbidden2} are not induced and $U$ is not an involution module.
    Then by definition we get two cases:
    \begin{enumerate}
    \item $\exists w,x \in U$, $\exists k \in \{1,...,|C|\}$ such that $N^k_w(X \backslash U) \neq N^k_x(X \backslash U)$ and
      $\exists l \in \{1,...,|C|\}$ such that $N^l_w(X \backslash U) \neq N^{\mathcal{I}(l)}_x(X \backslash U)$. 
      Then $D_1 = N^k_w(X \backslash U) \bigtriangleup N^k_x(X \backslash U)$, $D_2 = N^l_w(X \backslash U) \bigtriangleup N^{\mathcal{I}(l)}_x(X \backslash U)$,
      $I_1 = N^k_w(X \backslash U) \cap N^k_x(X \backslash U)$ and $I_2 = N^l_w(X \backslash U) \cap N^{\mathcal{I}(l)}_x(X \backslash U)$.\\
      Let $d_1 \in D_1$ and $i_1 \in I_1$, w.l.o.g we have $d_1$ connected to $w$ with color $k$. The color of the edge between $d_1$ and $x$ is thus
      $\mathcal{I}(k)$ otherwise we get figure \ref{forbidden2} induced in $G$.
      Therefore, the edges between $i_1$ and $x$ and $i_1$ and $w$ being of color $k$ we get the figure \ref{forbidden1}
      induced in $G$, a contradiction.
    \item $\forall w,x \in U$, $\forall k \in \{1,...,|C|\}$ such that $N^k_w(X \backslash U) = N^k_x(X \backslash U)$ such that
      $\forall l \in \{1,...,|C|\}$ such that $N^l_w(X \backslash U) = N^{\mathcal{I}(l)}_x(X \backslash U)$. 
      The involution has no fix point, this is a contradiction which concludes the proof.

    \end{enumerate}

  \end{proof}
\label{forbidden_pat}
\end{prop}

This proposition leads to the four following lemmas.

\begin{lem}{
    Let $G=(X,E)$ be a 2-structure, $\mathcal{I}$ an involution of the colors and
    $U$ and $V$ two crossing involution modules of $G$. 
    $U \cup V$ is an involution module of $G$.
  }
  \begin{proof}
    Assume towards contradiction that $U \cup V$ is not an involution module of $G$. Then by lemma \ref{forbidden_pat} 
    it has to contain an induced forbidden pattern. 

    If it contains the figure \ref{forbidden1}, then remark that $u,v \in U$ or $u,v \in V$ is not possible because $U$ and $V$ are
    involution modules.
    Assume w.l.o.g $u \in U \setminus V$, $v \in V \setminus U$ and $a,b \in X \setminus (U \cup V)$,
    but then, since $U$ and $V$ are crossing, $\exists w \in U \cap V$ and so, if $w$ does not induced 
    a forbidden pattern with $u$ it induces a forbidden pattern with $v$, a contradiction.

    Now, if it contains the figure \ref{forbidden2}, then remark that $u,v \in U$ or $u,v \in V$ is not possible because $U$ and $V$ are
    involution modules.
    Assume w.l.o.g $u \in U \setminus V$, $v \in V \setminus U$ and $a \in X \setminus (U \cup V)$, 
    but then, $\exists w \in U \cap V$ and so if $w$ does not induce a forbidden pattern with $u$ it induces a forbidden pattern with $v$, 
    a contradiction which concludes the proof.
  \end{proof}
  \label{union}
\end{lem}

\begin{lem}{
    Let $G=(X,E)$ be a 2-structure, $\mathcal{I}$ an involution of the colors and $U$ 
    and $V$ two crossing involution modules of $G$. 
    $U \cap V$ is an involution module of $G$.}
  \begin{proof}
    Assume towards contradiction that $U \cap V$ induces a forbidden pattern.
    
    First, it can not induce figure \ref{forbidden2} otherwise it will contradict the fact that $U$ and $V$ are involution modules.
    Now, if it induces a forbidden figure \ref{forbidden1} then $u,v \in U \cap V$ and w.l.o.g $a \in U \setminus V$ and $b \in V \setminus U$.
    But then, since $U$ and $V$ are crossing, there exists an element $w \in X \setminus (U \cup V)$ and so either $u,v,a,w$ or $u,v,b,w$ is a forbidden pattern.
    Because both $U$ and $V$ are involution modules, this is a contradiction which concludes the proof.

  \end{proof}
  \label{inter}
\end{lem}

\begin{lem}{
    Let $G=(X,E)$ be a 2-structure, $\mathcal{I}$ an involution of the colors
    and $U$ and $V$ two crossing involution modules of $G$. 
    $U \setminus V$ is an involution module of $G$.}
  \begin{proof}
    Assume towards contradiction that $U \setminus V$ is not an involution module of $G$.
    By lemma \ref{forbidden_pat} it induces a forbidden pattern.

    If it induces figure \ref{forbidden1}, then $u,v$ must be in $U \setminus V$ and we distinguish three cases for $a$ and $b$.
    Either $a,b \in V$; or $a \in U \cap V$ and $b \in X \setminus (U \cup V)$ or the other way around, $b \in U \cap V$ and $a \in X \setminus (U \cup V)$ 
    (the others cases induce a forbidden pattern for $U$).
    If $a,b \in V$, then $u,v,a,b$ induce a forbidden pattern for $V$ a contradiction.

    If $a \in U \cap V$ and $b \in X \setminus (U \cup V)$. Let $j = E(u,b)$ and $i = E(u,a)$ (it implies $E(v,b) = \mathcal{I}(j)$ and $E(v,a) = i$).
    Then, since $U$ and $V$ are crossing, it exists $w \in V \setminus U$.
    Let $k = E(u,w)$ and thus $E(v,w) = \mathcal{I}(k)$ (otherwise it induces a forbidden pattern for $U$).
    But then, $E(u,a) = E(u,b)$ and $E(u,w) = \mathcal{I}(E(v,w))$, it is a forbidden pattern for $V$, a contradiction. 

    Let us now prove the third case, if $b \in U \cap V$ and $a \in X \setminus (U \cup V)$. 
    Let $j = E(u,b)$ and $i = E(u,a)$ (it implies $E(v,b) = j$ and $E(v,a) = \mathcal{I}(i)$).
    Then, since $U$ and $V$ are crossing, it exists $w \in V \setminus U$.
    Let $k = E(u,w)$ and thus $E(v,w) = (k)$ (otherwise it induces a forbidden pattern for $U$).
    But then, $E(u,a) = \mathcal{I}(E(u,b))$ and $E(u,w) = E(v,w)$, it is a forbidden pattern for $V$, a contradiction.

    We now assume that $U \setminus V$ induces the figure \ref{forbidden2}.
    Then $u,v \in U \setminus V$ and necessarily $a \in U \cap V$.
    Since $U$ and $V$ are crossing, there exists $b \in V \setminus U$ and so, either $E(u,b) = E(v,b)$ or
    $E(u,b) = \mathcal{I}(E(v,b))$. In any case it induces a forbidden pattern with $a$, a contradiction which allows us to
    conclude the proof.

  \end{proof}
  \label{diff}
\end{lem}

\begin{lem}{
    Let $G=(X,E)$ be a 2-structure, $\mathcal{I}$ an involution of the colors
    and $U$ and $V$ two crossing involution modules of $G$. 
    $U \bigtriangleup V$ is an involution module of $G$.}

  \begin{proof}
    Assume towards contradiction that $U \bigtriangleup V$ is not an involution module of $G$.
    By proposition \ref{forbidden_pat} it induces a forbidden pattern.

    Assume that $U \bigtriangleup V$ induces figure \ref{forbidden1}. Then $u \in U$ and $v \in V$ 
    (otherwise it goes back to the case of lemma \ref{diff}).
    Now, we distinguish three different cases either $a,b \in U \cap V$ or $a \in U \cap V$ and $b \in X \setminus (U \cup V)$ or the other way around,
    $b \in U \cap V$ and $a \in X \setminus (U \cup V)$.
    
    We consider the first case. Since $U$ and $V$ are crossing there exists $w \in X \setminus (U \cup V)$ and then
    $a,b,w,u$ or $a,b,w,v$ induce a forbidden pattern for respectively $U$ or $V$, a contradiction.

    We now tackle the second case, namely $a \in U \cap V$ and $b \in X \setminus (U \cup V)$.
    Let $i = E(u,a) = E(v,a)$ and $j = E(u,b)$ (and thus $E(v,b) = \mathcal{I}(j)$).
    Then, since $U$ is an involution module, either $E(u,v) = i$ or $E(u,v) = \mathcal{I}(i)$.
    If $E(u,v) = i$ then $E(a,b) = j$ (otherwise it induces a forbidden pattern for $U$).
    This leads $u,v,a,b$ to be a forbidden pattern for $V$, a contradiction.
    If $E(u,v) = \mathcal{I}(i)$ then $E(a,b) = \mathcal{I}(j)$ (otherwise it induces a forbidden pattern for $U$).
    This also leads $u,v,a,b$ to be a forbidden pattern for $V$, a contradiction.

    We now address the third case, namely $b \in U \cap V$ and $a \in X \setminus (U \cup V)$.
    Let $i = E(u,a)$ and $j = E(u,b) = E(v,a)$ (and thus $E(v,a) = \mathcal{I}(i)$).
    Then, since $U$ is an involution module, either $E(u,v) = i$ or $E(u,v) = \mathcal{I}(i)$.
    If $E(u,v) = i$ then $E(a,b) = j$ (otherwise it induces a forbidden pattern for $V$).
    This leads $u,v,a,b$ to be a forbidden pattern for $U$, a contradiction.
    If $E(u,v) = \mathcal{I}(i)$ then $E(a,b) = \mathcal{I}(j)$ (otherwise it induces a forbidden pattern for $V$).
    This also leads $u,v,a,b$ to be a forbidden pattern for $U$, a contradiction.
    
    Let us assume that $U \bigtriangleup V$ induces figure \ref{forbidden2}. Then $u \in U$ and $v \in V$ 
    (otherwise it goes back to the case of lemma \ref{diff}) and $a \in U \cap V$.    
    Since $U$ is an involution module, either $E(u,v) = E(a,v)$ or $E(u,v) = \mathcal{I}(E(a,v))$.
    In any case, this induces a forbidden pattern for $U$ or for $V$, a contradiction.

    We conclude that $U \bigtriangleup V$ is an involution module of $G$.

  \end{proof}
\label{sym}
\end{lem}

These lemmas lead to the following theorem.

\begin{thm}{\textbf{Linear-sized tree representation.} The family of involution modules of any 2-structure is a partitive crossing 
    family and thus has a unique linear-sized tree-decomposition.}
\begin{proof}
By lemmas \ref{union}, \ref{inter}, \ref{diff} and \ref{sym} the family is closed under 
crossing union, intersection, difference and symmetric difference of its crossing members 
and so, it is a partitive crossing family.
Therefore by theorem \ref{decomp_theorem} the family has a unique linear-sized tree-decomposition.
\end{proof}
\end{thm}

\subsection{Tree-Decomposition Algorithm}

In this section, we present an $\mathcal{O}(n^2)$ algorithm which computes the tree representation of a family of 
involution modules of a 2-structure.

We first give an algorithm which computes the shape of the tree and the label of the nodes. We explain at the 
end of the section how to proceed in order to obtain the direction of the edges.
This means that we compute the tree-representation of not only the family of involution modules but the family of 
involution modules and their complement.

Before going into the details, let us first provide some intuition about the algorithm.
The idea is to modify the 2-structure in such a way that the tree-representation of the family of modules
of the new 2-structure has the same shape and same labels than the tree-representation of the family of involution modules
of the original 2-structure.
We first present how to modify the 2-structure and prove the properties of the transformation. 

In order to do so, for a given involution, we define a ternary operator on the colors of the edges of the 2-structure.

\begin{defn}{\textbf{Switch Colors.}}
  Let $G=(X,E)$ be a 2-structure, $C = \{1,...,c\}$ be the set of the colors of the 2-structure and $\mathcal{I}$ an involution of the colors.\\

  Let $C' = C$ $\cup$ $\{\Delta_{1,1}, ..., \Delta_{|C|,|C|}\}$ $\cup$ $\{\Delta_{1,1,1},...,\Delta_{|C|,|C|,|C|} \}$, 
  where the sets $\{\Delta_{1,1}, ..., \Delta_{|C|,|C|}\}$ and $\{\Delta_{\{1,2\},1},...,\Delta_{\{|C|-1,|C|\},|C|} \}$ 
  contain only new colors.\\

  We define the Switch\_Colors operator $\odot : C^3 \rightarrow C'$.\\

  $\forall i,j,k \in C$,
  \begin{itemize}
  \item $\odot(i,i,j) = \odot(i,\mathcal{I}(i),\mathcal{I}(j)) =  \Delta_{i,j} = \Delta_{\mathcal{I}(i),j}$, with 
    $i \neq j, \mathcal{I}(j)$;
  \item $\odot(i,j,k) = \Delta_{\{i,j\},k} = \Delta_{\{\mathcal{I}(i),j\},\mathcal{I}(k)} = \Delta_{\{i,\mathcal{I}(j)\},\mathcal{I}(k)} = \Delta_{\{\mathcal{I}(i),\mathcal{I}(j)\},k}$,
    with $k \neq \mathcal{I}(i),i,j,\mathcal{I}(j)$ and $i \neq j, \mathcal{I}(j)$;
  \item $\odot(i,j,i) = j$;
  \item $\odot(i,j,\mathcal{I}(i)) = \mathcal{I}(j)$.
  \end{itemize}
\end{defn}

\begin{figure}
\begin{center}

\includegraphics[scale=0.4]{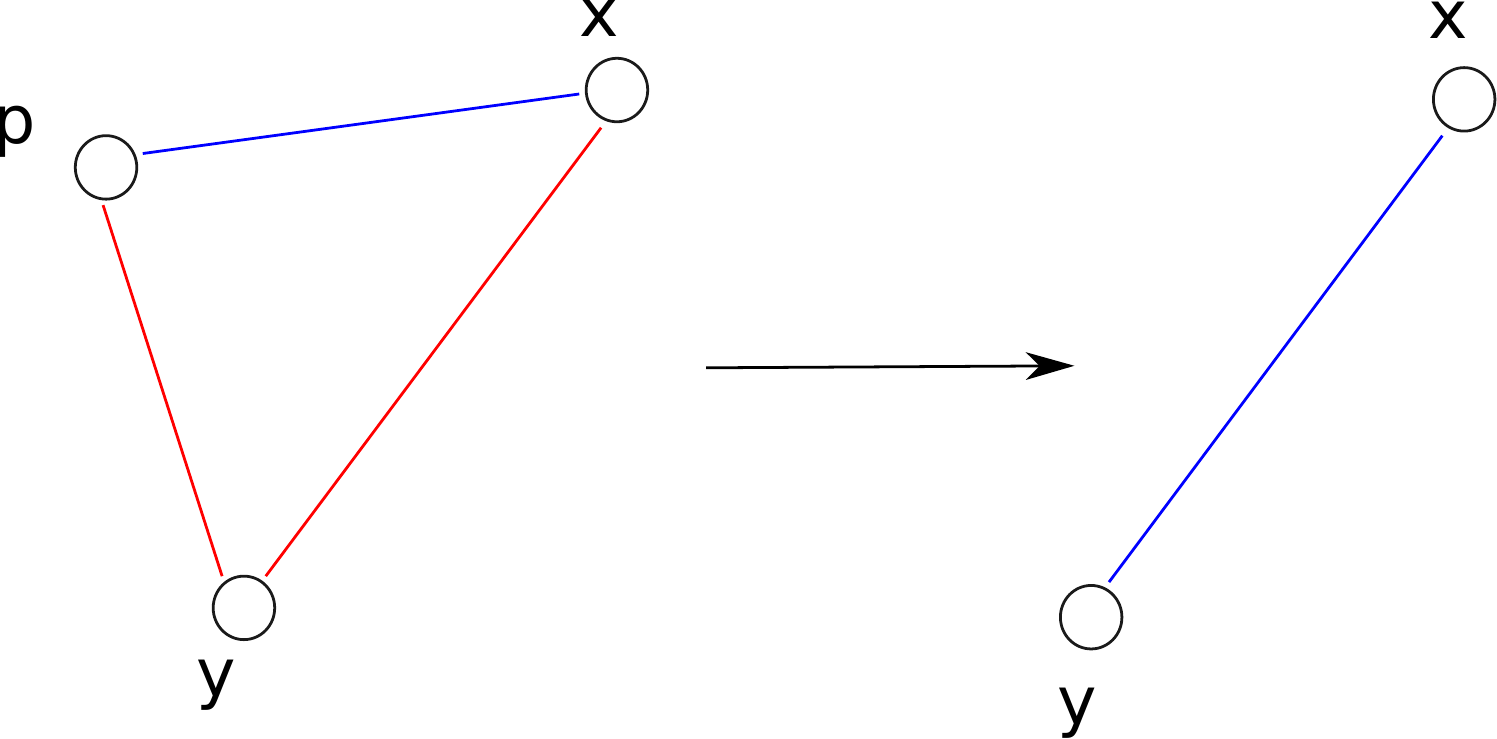}  

\caption{
  On the left a 2-structure $G$ with 3 vertices, on the right $G^{\odot}_p$.
}
\label{operator_example}

\end{center}
\end{figure}

Figure \ref{operator_example} illustrates how we apply the operator \emph{Switch Colors}.\\

\begin{defn}{\textbf{Switch Colors on 2-structures.}
    Let $G=(X,E)$ be a 2-structure, $\mathcal{I}$ an involution of its colors 
    and $s \in X$.
    We define $G^{\odot}_{s}$ as the 2-structure $(X',E')$, such that $X' = X \setminus \{s\}$ and 
    $\forall u,v \in X'$, $E'(u,v) = E'(v,u) = \odot (E(s,u),E(s,v),E(u,v))$.
}
\end{defn}

We mean here that we pick a vertex $s$, we call it the pivot, and for each couple of vertices $x,y$ different from $s$, we change the color
of the edge $\{x,y\}$ following the colors of the edge $\{s,x\}$, $\{s,y\}$ and $\{x,y\}$.
We now introduce the three following lemmas which ensure the correctness of the algorithm.

\begin{lem}{
    Let $G = (X,E)$ be a 2-structure with $c$ colors, $s \in X$, $G' = G^{\odot}_s = (X',E')$ and $\mathcal{I}$ an involution of the colors.\\
    Let $U \subset X$ such that $s \in U$, $U$ is an involution module of $G$. $X \setminus U$ is a module of $G'$.}
  \label{or1}
  \begin{proof}
    Let $U \subseteq X$ be an involution module of $G$ such that $s \in U$ and $M = X \setminus U$.
    Let $A$ be the set of elements of $U$ that have the same outside neighborhood than $s$ and $B = U \setminus A$.
    
    Then, for all elements $x \in A$ and $v \in M$, $E(x,v) = E(s,v)$.
    Therefore, when we apply the \emph{Switch\_Colors}, for any $x \in A$,
    we obtain $E'(x,v) = \odot (E(s,v), E(s,x), E(x,v)) = E(s,x)$.
    Thus, $x$ does not distinguish any element of $M$.

    Now, for all elements $x \in B$ and $v \in M$, $E(x,v) = \mathcal{I}(E(s,v))$.
    Therefore, when we apply the \emph{Switch\_Colors}, for any $x \in B$,
    we obtain $E'(x,v) = \odot (E(s,v), E(s,x), E(x,v)) = \mathcal{I}(E(s,x))$.
    Thus, $x$ does not distinguish any element of $M$.

    We can conclude that $M$ is a module of $G'$.
  \end{proof}
\end{lem}

\begin{lem}{
    Let $G = (X,E)$ be a 2-structure with $c$ colors, $s \in X$, $G' = G^{\odot}_s$ 
    Let $U \subset X$ such that $s \in U$, $X \setminus U$ an involution module of $G$. $X \setminus U$ is a module of $G'$.}
  \label{or2}
  \begin{proof}
    Let $U \subseteq X$ and $M = X \setminus U$ such that $s \in U$ and $M$ is an involution module.
    Then $s$ partitions $M$ into 2 parts $A$ and $B$ such that for any elements $a \in A$ and $b \in B$, 
    $E(s,a) = \mathcal{I}(E(s,b))$.

    Let $x$ be an element of $U$. Note that $x$ also splits $M$ into the same parts $A$ and $B$ (otherwise the elements of $M$ do not
    partition the graph the same way).
    When we apply the \emph{Switch\_Colors} on $G$, for all $a \in A$, 
    $E'(x,a) = \odot (E(s,a), E(s,x), E(x,a))$ and for all $b \in B$,
    $E'(x,b) = \odot (E(s,b), E(s,x), E(x,b)) = \odot (\mathcal{I}(E(s,a)), E(s,x), \mathcal{I}(E(x,a)))$.
    Hence, according to the definition of the \emph{Switch\_Colors}, 
    $\odot (\mathcal{I}(E(s,a)), E(s,x), \mathcal{I}(E(x,a))) =  \odot (E(s,a), E(s,x), E(x,a))$ and 
    therefore $E'(x,a) = E'(x,b)$.
    $x$ does not distinguish any element of $M$ in $G'$, we conclude that $M$ is a module of $G'$.
  \end{proof}
\end{lem}

We now prove the converse of the two previous lemmas.

\begin{lem}{
    Let $G = (X,E)$ be a 2-structure with $c$ colors, $s \in X$ and $G' = G^{\odot}_s$ and $\mathcal{I}$ an involution of the colors.\\
    Let $U \subset X$ such that $s \in U$, $M = X \setminus U$ is a module of $G'$. $U$ is an involution module of $G$ or
    $X \setminus U$ is an involution module of $G$.}
  \label{recip}
  \begin{proof}
    
    Assume towards contradiction that $M$ is a module of $G'$ and neither $U$ nor $M$ are involution modules of $G$.
    Since the singletons are involution modules, note that $|M| > 1$ and $|U| > 1$.

    If $U$ and $M$ are not involution modules then each of them induce a forbidden pattern in $G$.
    
    Assume first that $M$ induces figure \ref{forbidden1}.
    Then we pick $w \in U$ and $u,v \in M$ such that $s,w,u,v$ induces figure \ref{forbidden1}.
    If $M$ is a module of $G'=(V',E')$ then $\odot (E(s,w),E(s,u),E(w,u)) = \odot (E(s,w),E(s,v),E(w,v))$.
    There are only two possible cases, either $E(s,u) = E(s,v)$ and $E(w,u) = \mathcal{I}(E(w,v)) $ or 
    $E(s,u) = \mathcal{I}(E(s,v))$ and $E(w,u) = E(w,v)$.
    Hence, by definition of \emph{Switch\_Colors}, in any case $\odot (E(s,w),E(s,u),E(w,u)) \neq \odot (E(s,w),E(s,v),E(w,v))$,
    a contradiction.

    Assume now $M$ induces figure \ref{forbidden2}. We can pick $v \in U$ and $a,b \in M$ such that $a,b,v$ 
    induce figure \ref{forbidden2} then.
    Now we distinguish the two possible cases, either $U$ is not an involution module because it induces figure \ref{forbidden2}
    or because it induces figure \ref{forbidden1}.

    In the first case, we can pick $u \in U$ and $a \in M$ such that $s,u,a$ induce figure \ref{forbidden2}.
    Now, since $M$ is a module of $G'$, we have $E'(v,a) = E'(v,b)$ and $E'(u,a) = E'(u,b)$.
    Hence, $\odot (E(s,a), E(s,v), E(v,a)) = \odot (E(s,b), E(s,v), E(v,b))$ and $E(v,b) \neq E(v,a), \mathcal{I}(E(v,a))$.
    By the \emph{Switch\_Colors} definition, this is true if and only if $E(s,a) = E(v,a)$ and $E(s,b) = E(v,b)$
    or $E(s,a) = \mathcal{I}(E(v,a))$ and $E(s,b) = \mathcal{I}(E(v,b))$. 
    Now, since $M$ is a module of $G'$, $\odot (E(s,a), E(s,u), E(u,a)) = \odot (E(s,b), E(s,u), E(u,b))$ and 
    $E(s,a) \neq E(u,a), \mathcal{I}(E(u,a))$.
    By the \emph{Switch\_Colors} definition, this is true if and only if $E(s,a) = E(s,b)$ or $E(s,a) = \mathcal{I}(E(s,b))$,
    a contradiction.

    In the latter case, we pick $b,c \in M$ and $u \in U$ such that $u,s,b,c$ induce figure \ref{forbidden1}.
    Now, either $E(s,b) = E(u,b)$ and $E(s,c) = \mathcal{I}(E(u,c))$ or $E(s,b) = \mathcal{I}(E(u,b)$ and
    $E(s,c) = E(u,c)$.
    If $E(s,b) = E(u,b)$ and $E(s,c) = \mathcal{I}(E(u,c))$ then $E'(u,b) = \odot (E(s,b), E(s,u), E(u,b)) = E(s,u)$ 
    and $E'(u,c) = \odot (E(s,c), E(s,u), E(u,c)) = \mathcal{I}(E(s,u))$. Therefore $u$ distinguishes $c$ from $b$.
    $M$ is not a module of $G'$, a contradiction.\\
    If $E(s,b) = \mathcal{I}(E(u,b)$ and $E(s,c) = E(u,c)$ then $E'(u,b) = \odot (E(s,b), E(s,u), E(u,b)) = \mathcal{I}(E(s,u))$ 
    and $E'(u,c) = \odot (E(s,c), E(s,u), E(u,c)) = (E(s,u)$. Therefore $u$ distinguishes $c$ from $b$.
    $M$ is not a module of $G'$, a contradiction which concludes the proof.
  \end{proof}
\end{lem}

These three lemmas induce the following theorem.

\begin{thm}{ 
    Let $G = (X,E)$ be a 2-structure with $c$ colors, $s \in X$ and $G' = G^{\odot}_s$.\\
    Let $U \subset X$ such that $s \in U$. $M = X \setminus U$ is a module of $G'$ $\iff$ $U$ is an involution module of $G$ or
    $X \setminus U$ is an involution module of $G$.}
  \label{thm_decomp_switch1}
\end{thm}

This theorem is of particular importance because it guarantees that the tree-representation of the family of involution modules
of any 2-structure $G=(X,E)$ is \emph{almost} the same than the tree-representation of the family of modules of the 2-structure
$G^{\odot}_s$, for any $s \in X$. 
This is what we state below.

\begin{prop}{
    Let $G=(X,E)$ be a 2-structure and $s$ be an element of $X$. 
    The involution modular decomposition tree $\mathcal{T}$ of $G$ and the modular decomposition tree
    $\mathcal{T}_{G^{\odot}_s}$ of $G^{\odot}_s$ have the following properties:
      \begin{itemize}
      \item The two trees have the same nodes except that the leaf with label $s$ is missing
        in $\mathcal{T}_{G^{\odot}_s}$ but present in $\mathcal{T}$.
        \item The node of $\mathcal{T}$ that is adjacent to the leaf $s$ corresponds to the root of $\mathcal{T}_{G^{\odot}_s}$ 
          (while $\mathcal{T}$ is unrooted).
        \item The prime and complete nodes are the same in both trees.
      \end{itemize}
  }
  \begin{proof}
    This is a direct consequence of theorem \ref{thm_decomp_switch1}. Each strong module of $\mathcal{T}_{G^{\odot}_s}$ is a strong
    involution module or the complement of a strong involution module of $G$ and the converse holds.
    Therefore, for any complete node $N$ of $\mathcal{T}_{G^{\odot}_s}$, the union of any subset of the neighbors of $N$ is a 
    module of $G^{\odot}_s$ and thus it is an involution module or the complement of an involution module of $G$.
    The same reasoning applies for the prime nodes.
    For each involution module $U$ and its complement $X \setminus U$, the part which contains $s$ is dropped and the other
    part is included in the family of modules of $G^{\odot}_s$. Thus, the neighbor of node $s$ in $\mathcal{T}_{G}$ is the root
    of $\mathcal{T}_{G^{\odot}_s}$.
  \end{proof}
  \label{prop}
\end{prop}

For any 2-structure, the tree computed by our algorithm is exactly an undirected version of the 
tree-representation of the family of involution modules of the 2-structure. 

We now show how to determine the direction of the edges of the tree.
Let us first recall a theorem from \cite{BHR}.

\begin{thm}{\cite{BHR}. The tree-representation $\mathcal{T}$ of any weakly partitive crossing family has either one sink or 
    only one double-arc $uv$ such that $\mathcal{T} \setminus uv$ has two sinks $u$ and $v$.
  }
  \label{thm_direct}
\end{thm}

We now proceed bottom-up in order to direct the edges.
The algorithm is as follow, first we direct the edges until we find a vertex which has only in arcs.
By theorem \ref{thm_direct}, either this vertex is the sink of the tree or it shares a double arc with one of its neighbors. 
We then consider the edges that are adjacent to the sink vertex in order to determine whether there is a double arc or not.

\begin{defn}{\textbf{Edge Direction Algorithm}.}
\paragraph{Phase 1}
We begin by the leaves - which are always involution modules so that they all have an out arc. Then for each leaf $l$
we can check whether $X \setminus \{ l \}$ is an involution module. 
If we find a leaf whose complement is also an involution module then we are done: the out arc of the leaf is the double-arc and we
direct the edges to the leaf.

\paragraph{Phase 2}
Then we perform bottom-up by considering all the nodes that have at most one edge undirected.
If a node has one out arc then we direct the other edges to the node.
Now, consider a node $N$ with $k+1$ neighbors with only one undirected edge and $k$ in arcs.
For each neighbor $V$, we pick a vertex which is a leaf of the subtree rooted at $V$.
Call the set of chosen vertices $S$ and let $W$ be the neighbor whose edge to $N$ is undirected.
Then, we check that $S$ is an involution module for the 2-structure $G_{[S \cup V(W)]}$ where $V(W)$ is 
the set of leaves whose paths to $N$ traverse $W$.
If this set is an involution module then the union of the sets of the leaves of the subtrees rooted at the processed
neighbors of $V$ is an involution module (since each set is an involution module and because of the union stability).
We can therefore direct the edge from $V$ to $W$.
Otherwise we direct the edge from $W$ to $V$.
Phase 2 terminates when we find a sink vertex $u$.

\paragraph{Phase 3}
Now, we only need to test whether this sink vertex has a double arc with one of its neighbor.
Note that for each neighbor $N$ the set of leaves of the subtree rooted at $N$ is an involution module 
of $G$.
Let $k$ be the number of neighbors of $u$ and $N_k$ be the set of leaves of the subtree rooted at the $k^{th}$
neighbor of $u$.
For each of the $k-1$ remaining neighbors, we pick a vertex. Let call $S$ the set of these vertices.
We first check whether this set is an involution module of the 2-structure $G_{[N_1 \cup S]}$.
If it is, then we are done.
Otherwise we drop the vertex of the second neighbor and we pick a vertex of the first neighbor and we check
whether this set is an involution module of the 2-structure $G_{[N_2 \cup S]}$ and so on until we find
a set which is an involution module (and thus a double arc) or not (and thus $N$ is the unique sink of the tree).

\end{defn}

We can now state the following theorem.

\begin{thm}{The Edge Direction Algorithm computes the direction of the edges of the involution modular tree-decomposition of 
    any 2-structure $G=(X,E)$ with an $\mathcal{O}(|X|^2)$ time complexity.}
  \label{edge_direction_algo_thm}
  \begin{proof}
    The correctness of the algorithm follows from theorem \ref{thm_direct}, lemma \ref{union} and the definition.
    
    We now show that the time complexity of the algorithm is $\mathcal{O}(|X|^2)$.
    Notice first that one can greedily check whether a set of size $k$ is an involution module of a 2-structure $G=(X',E')$ in 
    $C.k.(|X'| - k)$ operations for some constant $C$ by checking for each vertex of the set if the partition of the rest of the 2-structure coincides
    with the partition of the already processed vertices and by reccording an adjacency matrix of the colors of the 2-structure.
    
    The cost of phase 1 is thus $\mathcal{O}(|X|^2)$ since there is exactly $|X|$ leaves.
    
    Then during phase 2, for each node $N$ the cost is at most $C.k.|X|$ where $k$ is the number of neighbors of $N$.
    By taking the sum over all the nodes of the tree we obtain an $\mathcal{O}(|X|^2)$ for the complexity of phase 2.

    Let us now consider the third phase.
    Assume that $U$ has $k$ neighbors.
    For the $i^{th}$ neighbor we have to check whether the set $S$ is an involution module of the 2-structure $G_{[S \cup N_i]}$.
    The cost is at most $C.k.|N_i|$.
    Note that the sum of all the $N_i$ is exactly $|X|$.
    Therefore by taking the sum over all the neighbors of $U$ we obtain an overall cost of $C.k.|X|$.
    Since $k \le |X|$, the complexity of phase 3 is $\mathcal{O}(|X|^2)$.
    
    Therefore, the complexity of the algorithm is $\mathcal{O}(|X|^2)$.
        
  \end{proof}
\end{thm}

Theorem \ref{edge_direction_algo_thm} and proposition\ref{prop} allow us to conclude this section with the following theorem.

\begin{thm}{
    There exists an $\mathcal{O}(n^2)$ algorithm which computes the tree representation of the family
    of the involution modules of any 2-structure.}
  \begin{proof}
    The algorithm consists in picking a vertex $s$ and
    applying the operator \emph{Switch Colors} to the 2-structure. This can be done 
    in $\mathcal{O}(n^2)$ by considering each edge once and applying the rules described above.
    Then we apply the $\mathcal{O}(n^2)$ modular decomposition algorithm of \cite{ehrenfeucht} 
    and we obtain the tree.
    We then apply the Edge Direction Algorithm in order to compute the direction of the edges of the tree.
    
    Theorems \ref{thm_decomp_switch1} and \ref{edge_direction_algo_thm} and proposition \ref{prop} ensure the correctness of the algorithm.    
  \end{proof}
\end{thm}

Before moving to the next section, we recall the definition of the Seidel Switch and remark that our Switch Colors operator generalizes the Seidel Switch
to 2-structures.

\begin{defn}{\textbf{Seidel Switch}. Let $G = (X,E)$ be a graph and $v \in E$. The Seidel Switch applied at $v$ on $G$ consists in complementing the edges and
non-edges of neighbors and non-neighbors of $v$ before removing $v$. The resulting graph is
\begin{center}
  $G' = (X \setminus v, E')$ where $E' = E \bigtriangleup \{xy | vx \in E, vy \notin E \}$.
\end{center}}

\end{defn}

\begin{rem}{\textbf{Seidel Switch}. The \emph{Switch Colors} operator applied to undirected graphs
    coincides with the Seidel Switch defined in \cite{Seidel_Switch}. One can see the \emph{Switch Colors} operator
    as a generalization of the Seidel Switch to 2-structures.
  } 
\end{rem}

\section{Switch Cographs}
\label{switch_cographs}
We now focus on undirected graphs - which are symmetric 2-structures with two colors - and we use our new decomposition tool in order
to state structural properties and design algorithms.
Let us first remark that there is only one involution without fix point for the case of graphs so that we do not have
to quantify on the involution throughout this section.


The modular decomposition led to study the classes of graphs which have a particular tree-decomposition. The best-known
class is the class of cographs whose tree-decompositions have only complete nodes (they are called \emph{completely
decomposable} with respect to modular decomposition). \cite{corneil_P4} that 
the class of cographs is exactly the class of $P_4$-free graphs (i.e the class with no induce path with four vertices). 
\cite{brandstadt} showed that the class of ($P_5$, Gem)-free graphs is a good generalization of the class of 
cographs since they have Clique-width at most 5 and thus some classical graph problems (the stable set problem for example) are polynomially
tractable. Nevertheless this class only provides a Clique-width decomposition expression and no tree-decomposition (unlike cographs).
Tree-decomposition is a powerful tool that can led to solve even more problems than a Clique-width decomposition expression 
(which helps to solve problems expressible by monadic second order logic without edge set quantification \cite{courcelle_msol}) for particular
classes of graphs.

Since the concept of involution module generalizes strictly the concept of module, an obvious and well-founded problem to 
address consists in characterizing and studying the class of graphs completely decomposable 
with respect to involution modular decomposition. This class of graphs generalizes strictly the class of cographs.

\begin{figure}
\begin{center}

\includegraphics[scale=0.4]{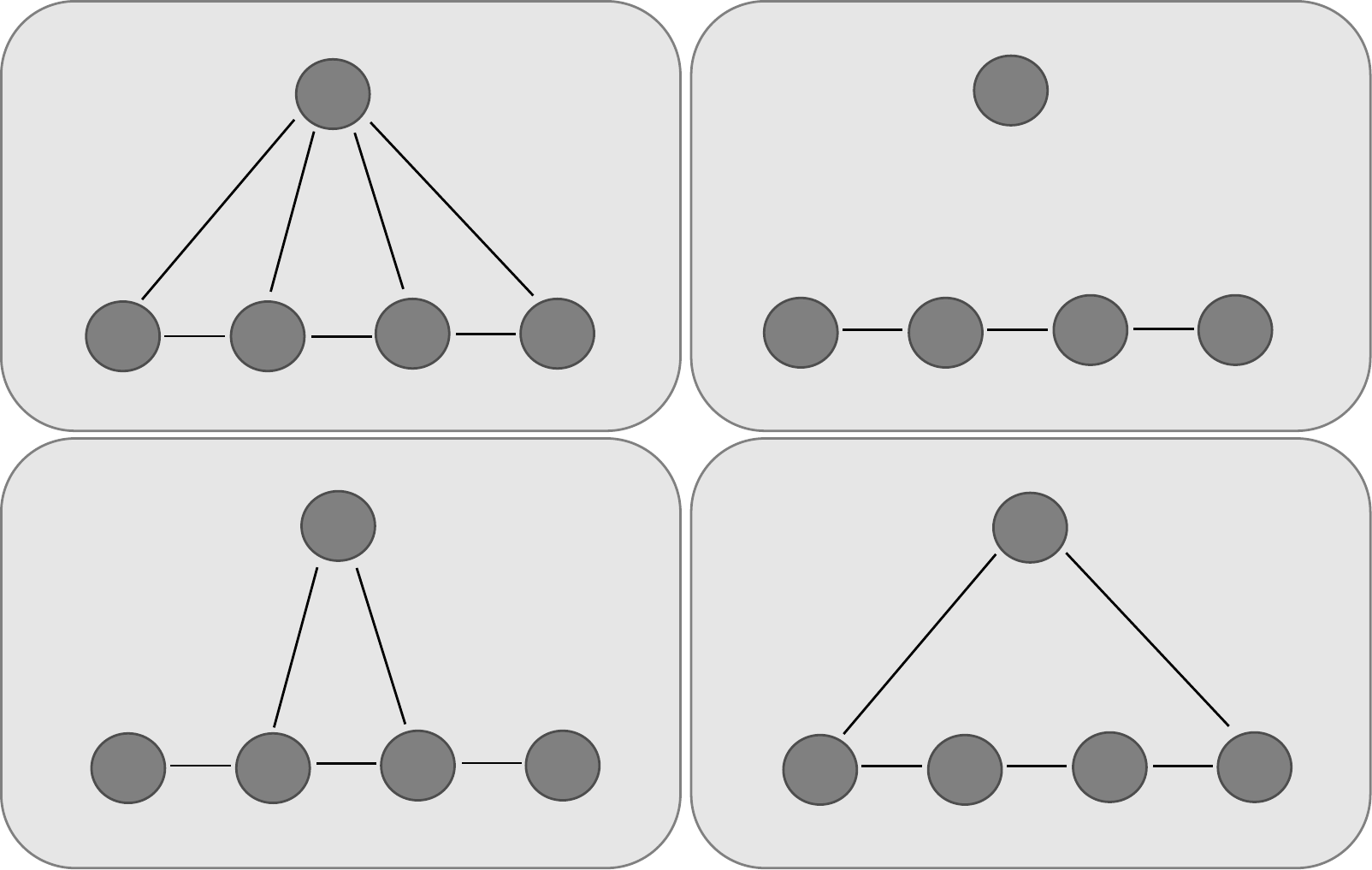}
\caption{From left to right and top to bottom, the Gem, Co-gem, Bull and $C_5$ graphs}
\label{forbidden_induced_subgraphs}
\end{center}
\end{figure}

We show that this class is the class of (Gem, Co-gem, Bull, $C_5$)-free graphs (refer to figure \ref{forbidden_induced_subgraphs}), 
introduced by \cite{hertz} as the class of switching-perfect graphs - the class of graphs which leads to a perfect graph after a Seidel Switch - 
and studied by \cite{bijoin} who gave a linear algorithm for the switch cograph isomorphism problem.
We use the involution modular decomposition to tackle well-known graph problems.

First, we begin by highlighting structural properties of particular importance.

\begin{defn}{\textbf{(Twin,Antitwin)-extension}.} A (twin,Antitwin)-extension of a graph $G$ is a graph $G'$ which consists of $G$ and a new vertex $v$ which
is either a twin (i.e it has the same neighborhood than another vertex) or an antitwin (i.e the complement of its neighborhood coincides with the neighborhood of another vertex)  
of at least one vertex of $G$.
\end{defn}

\subsection{Structural Properties}
\begin{thm}{Let $G$ be an undirect graph. The following definitions are equivalent:
    \begin{enumerate}
    \item $G$ is a switch cograph;
    \item The umodular and involution modular decomposition trees of $G$ do not contain any prime node;
    \item Let $p \in V$, and $G'$ the graph corresponding to $G$ after a Seidel Switch on $p$.
      $G'$ is a cograph;
    \item $G$ has no induced Gem, Co-Gem, $C_5$ nor Bull subgraphs;
    \item The class of switch cographs is the class of graphs which can be obtained from a single vertex by a sequence of (twin,antitwin)-extensions.
    \end{enumerate}
  }

  \begin{proof}
    \cite{bijoin} showed that $1 \iff 3 \iff 4$. \\
    Theorem \ref{thm_decomp_switch1} implies that $1 \iff 2$ because the labels of the two trees are the same.
\end{proof}

\end{thm}

\begin{lem}{Binary Decomposition Tree.}
For any Switch Cograph, there exists a decomposition tree with maximum degree equal to 3.
\begin{proof}
Let $G = (V,E)$ be a switch cograph, $p \in V$, and $G'$ the graph corresponding to $G$ after a Seidel Switch on $p$.
Since the decomposition tree of $G$ coincides with the decomposition tree of $G'$, $G$ has maximum degree equal to 3 if
and only if $G'$ has maximum degree equal to 3.
There exists a lemma from \cite{cotree}, saying that any cograph has a tree representation with degree at most 3 which concludes the proof.
\end{proof}
\end{lem}

Notice that this tree is not canonical.
Throughout this section, for any graph $G=(V,E)$ we denote by $n$ the cardinality of set $V$ and by $m$ the cardinality of set $E$.

\begin{rem}{The class of switch cographs is closed under complement because the set of forbidden subgraphs is closed under complement
    Besides, the decomposition tree of the complement graph of any switch cograph can be computed in $\mathcal{O}(n+m)$ by computing the involution modular decomposition tree
    and changing each clique node into a bipartite node and vice versa.}
\end{rem}

\begin{lem}{Every switch cograph is a perfect graph.}
\begin{proof}
First notice that there is no hole nor anti-hole of length five since the switch cographs are $C_5$-free.
Then, if there is an odd hole (resp. an odd anti-hole) of lenght greater than 7, it contains
an induced Co-gem (resp. an induced Gem) which is a forbidden subgraph.
Thus, switch cographs are bull-free berge graphs and so, perfect by \cite{chvatal}.
\end{proof}
\end{lem}

\subsection{Algorithmic paradigm}
Throughout this section, we propose algorithms which traverse the decomposition tree of the switch cographs in a bottom-up fashion in the same
way as it is done in \cite{cotree} for the cographs.
Namely, for any switch cograph $G$ an edge of its binary involution modular decomposition tree is picked 
and an artificial node is created on it. Then the tree is rooted at this node.
A node is processed when its two children have already been processed. 

Let us now introduce some notations, for any switch cograph $G$ and for any node $N$ of its binary involution modular decomposition tree 
and its two children $A$ and $B$, we note $G_{[N]}$ the subgraph of $G$ induced by the leaves of the subtree rooted at $N$.

\begin{figure}
\begin{center}

\begin{multicols}{2}
\includegraphics[scale=0.5]{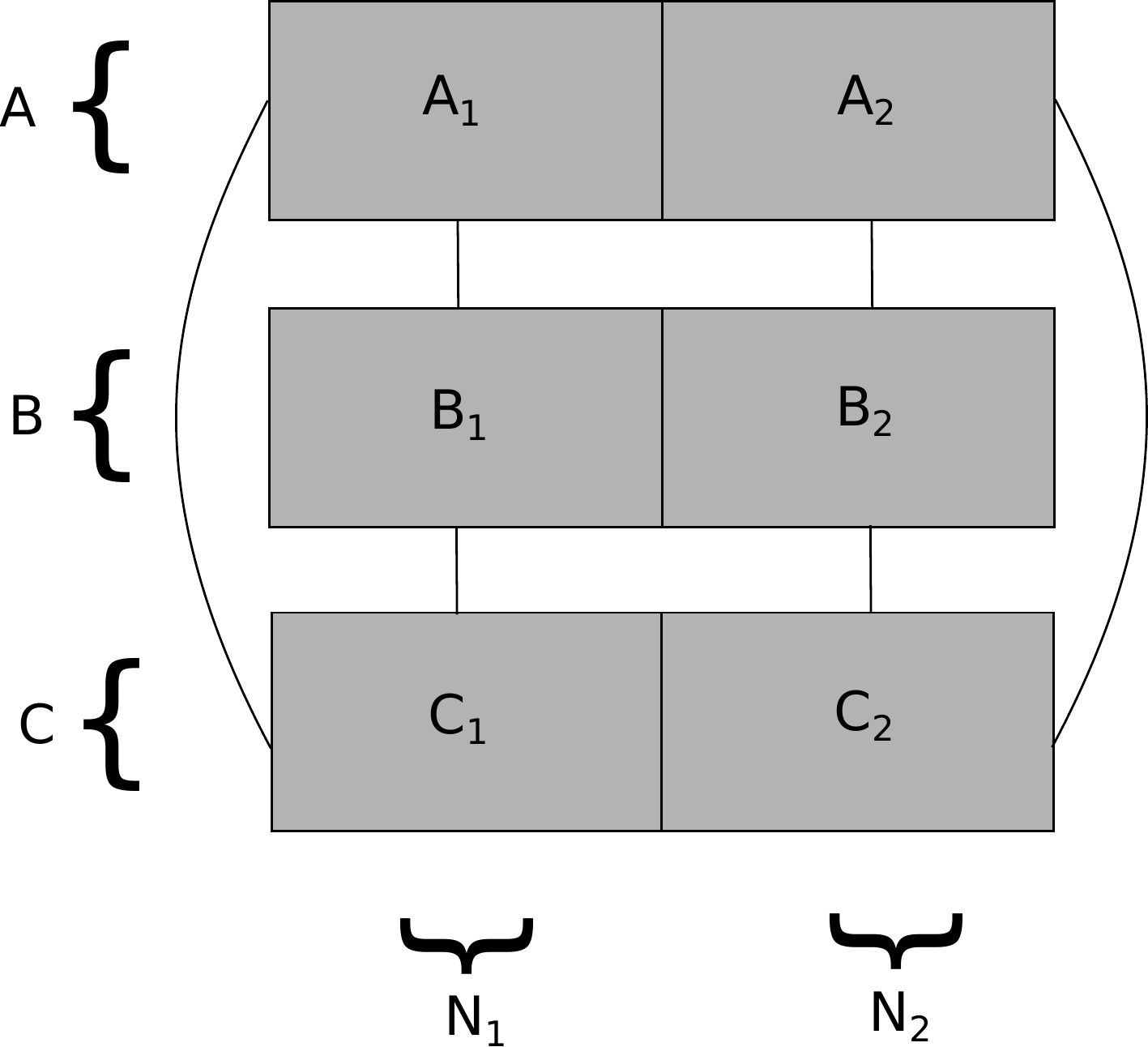}
\caption{\textbf{Clique Node}.
A node $N = (N_1,N_2)$ of a binary IMDT of a switch cograph such 
that $A=(A_1,A_2)$, $B=(B_1,B_2)$ and $C=(C_1,C_2)$ are its three neighbors and $A_1,B_1,C_1$ are connected in a 
clique fashion and $A_2,B_2,C_2$ as well.}
\label{node_tree}

\includegraphics[scale=0.5]{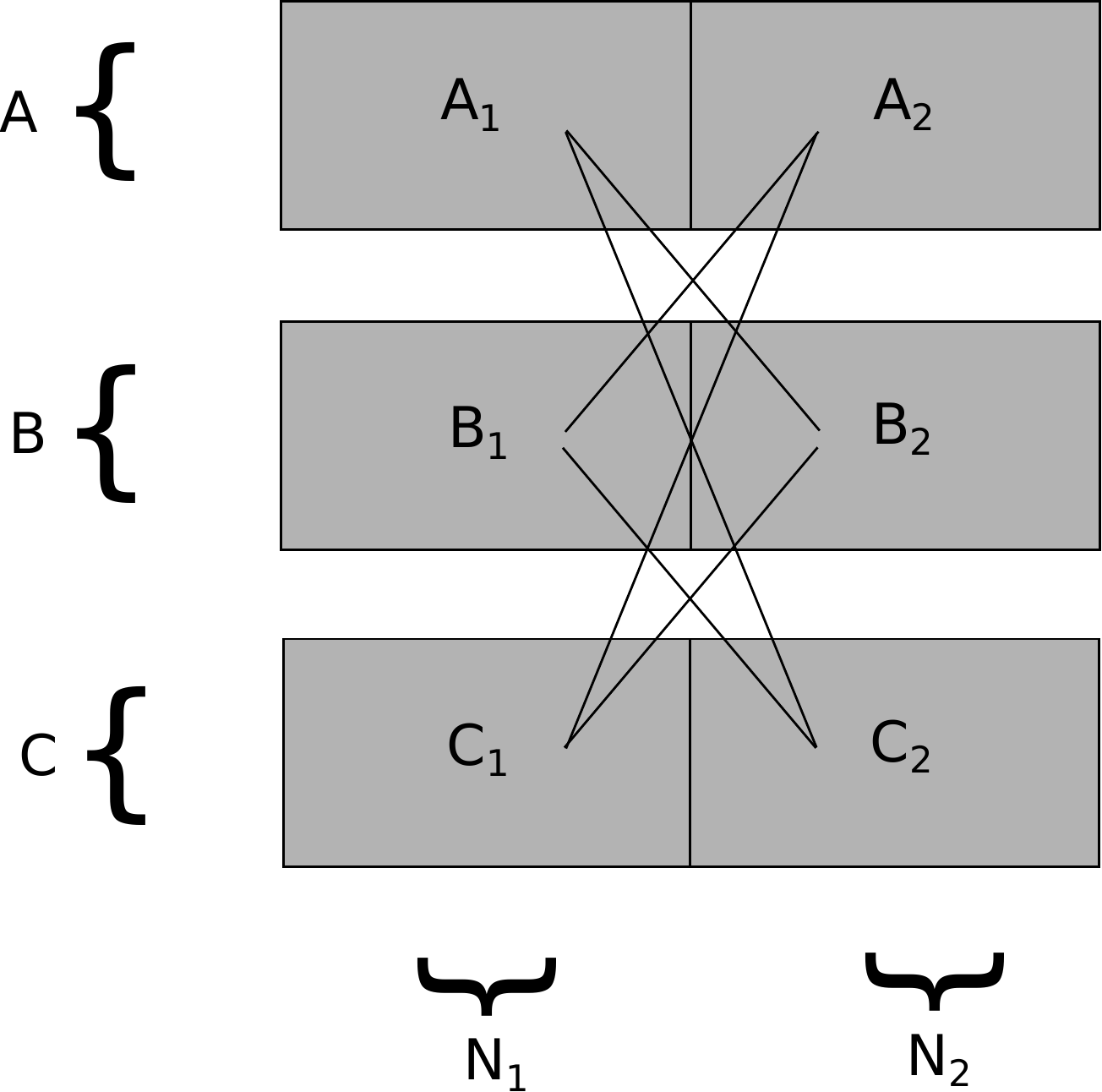}
\caption{\textbf{Bipartite Node}.
A node $N = (N_1,N_2)$ of a binary IMDT of a switch cograph such 
that $A=(A_1,A_2)$, $B=(B_1,B_2)$ and $C=(C_1,C_2)$ are its three neighbors and they are connected in a 
bipartite fashion.}
\label{node_tree_bip}

\end{multicols}

\end{center}

\end{figure}

By \cite{bijoin}, the nodes of the binary involution modular decomposition tree are of two kinds. 
We distinguish the clique node (figure \ref{node_tree}) and the bipartite node 
(figure \ref{node_tree_bip}). In any case, the graph $G_{[N]}$ can be split into two parts such that there exists $A_1,A_2$ bipartition of 
$A$; $B_1,B_2$ bipartition of $B$; and $C_1,C_2$ bipartition of $C$ (where $C$ is the parent of $N$ in the rooted tree) such that,
for the clique node, there are all the edges between the elements of $A_1$ and $B_1 \cup C_1$ and no edge to elements of $B_2 \cup C_2$, 
there are all the edges between elements of $A_2$ and elements of $B_2 \cup C_2$ and no edge to elements of $B_1 \cup C_1$, 
there are all the edges between elements of $B_1$ and elements of $A_1 \cup C_1$ and no edge to elements of $A_2 \cup C_2$ and 
there are all the edges between elements of $B_2$ and elements of $A_2 \cup C_2$ and no edge to elements of $A_1 \cup C_1$.
The bipartite node is the complement of the clique node, i.e builds the clique node and complements the edges and non-edges created.

We note $N=(N_1,N_2)$ to refer to the node $N$ and its two parts. 

This lead to the following crucial lemma.

\begin{lem}{Let $G=(V,E)$ be a switch cograph and $N = (N_1,N_2)$ a node of its involution modular binary decomposition tree and $A=(A_1,A_2)$ and $B=(B_1,B_2)$ 
    its two children. 
    Then, either $N_1 = A_1 \cup B_1$ and $N_2 = A_2 \cup B_2$; or $N_1 = A_1 \cup B_2$ and $N_2 = A_2 \cup B_1$; or 
    $N_1 = A_2 \cup B_2$ and $N_2 = A_1 \cup B_1$; or $N_1 = A_2 \cup B_1$ and $N_2 = A_1 \cup B_2$.
}

\label{crucial_lemma}
\begin{proof}
Assume towards contradiction that there exists a clique node $N = (N_1,N_2)$ with two children $A = (A_1,A_2)$ and $B =(B_1,B_2)$ such that 
the bipartitions of $A$ and $B$ is not respected at $N$. Let $C=(C_1,C_2)$ be the third neighbor of $N$.

Now, let $A_1' = N_1 \cap A$, $B_1' = N_1 \cap B$, $A_2'=N_2 \cap A$ and $B_2'= N_2\cap B$.
W.l.o.g we can assume that $A_1 \cap N_1 \neq \emptyset$ and $A_1 \cap N_2 \neq \emptyset$.
Consider now the node $A=(A_1,A_2)$, and its two children $D$ and $E$ and its third neighbor which is the rest of the graph, namely
$G_{[B \cup C]}$. 

Then there are all the edges between the elements of $C_1 \cup B_1'$ and the elements of $A_1'$.
So if $\exists a \in A_1' \setminus A_1$, it implies that $a$ is connected to every element of $C_1 \cup B_1'$ and so $a \in A_1$, 
a contradiction.
If $\exists a \in A_1 \setminus A_1'$, it implies that $a$ has no edges with $C_1 \cup B_1'$ and so $a \notin A_1$, a contradiction.
Thus, $A_1 = A_1'$, a contradiction.

The same reasoning applies to the bipartite node case.
\end{proof}
\end{lem}

Then for any node $N=(N_1,N_2)$, we refer to its two children $A$ and $B$ as $A=(A_1,A_2)$ and $B=(B_1,B_2)$ such that
$N_1 = A_1 \cup B_1$ and $N_2 = A_2 \cup B_2$.
Thus, for a clique node $A_1$ and $B_1$ are connected in a clique fashion and $A_2$ and $B_2$ as well and for
a bipartite node $A_1$ and $B_1$ are not adjacent and $A_2$ and $B_2$ either (see figures \ref{node_tree} and \ref{node_tree}).

\begin{lem}{Let $G=(V,E)$ be a switch cograph and $N=(N_1,N_2)$ a node of its involution modular binary decomposition tree (IMDT). 
    $G_{[N_1]}$ and $G_{[N_2]}$ are cographs.
  }
\begin{proof}
First, for all $i,j \in \{1,2\}$, $i\neq j$,
$G_{[A_i]}$ (resp. $G_{[B_i]}$) is a cograph (proof: if $A_i$ or $A_j$ (resp. $B_i$ or $B_j$) contains a $P_4$
and $B_i \cup B_j$ (resp. $A_i \cup A_j$) is not empty there is an induced Gem or Co-gem).
Then, since $A_i$ and $B_i$ are both cographs and are either connected in a clique fashion or not adjacent at all, 
by lemma \ref{crucial_lemma} $N_i$ is a cograph.
\end{proof}
\end{lem}

Since the IMDT and the umodular decomposition tree coincide for graphs, we recall the following
lemma.

\begin{lem}{\cite{umodules}. The binary IMDT of a switch cograph can be computed in $\mathcal{O}(n+m)$.
}
\label{binary_IMDT_complexity}
\end{lem}

\begin{rem}{
Notice that the IMDT provides an efficient tool for switch cographs recognition.
To test whether a graph is a switch cograph, compute its IMDT and check that
each node of the tree is a complete node. 
These operations can be done in $\mathcal{O}(n+m)$ thanks to lemma \ref{binary_IMDT_complexity}.}
\end{rem}

Before we start, we state the following lemma that will help 
for the complexity analysis of the following algorithms.

\begin{lem}{Let $\mathcal{T}$ a rooted binary tree of size $\mathcal{O}(n)$ and $\mathcal{A}$ be an algorithm which proceeds
    bottom-up on $\mathcal{T}$.

    Let $N$ be a node of $\mathcal{T}$ whose subtree contains $n_N$ nodes and $A$ and $B$ be its two children whose subtrees respectively
    contains $n_A$ and $n_B$ nodes. 
    If the running time of the algorithm at node $N$ assuming that its two children have already been computed is less than $n_A^{C/2} * n_B^{C/2}$,
    then the overall complexity of the algorithm is $\mathcal{O}(n^{C})$, for some constant $C \ge 1$.
}

\label{lem_complexity}
\begin{proof}
  First, notice that the complexity of the algorithm is some constant $c$ for the leaves of the tree.
  We show by induction that the complexity of the algorithm at a node $N$ is 
  less than $c . n_N^C$.
  We assume that this holds for any node at a distance of at most $k$ from a leaf.
  We show that this is true for a node at a distance  $k+1$.
  Let $N$ be such a node and $A$ and $B$ be its two children.
  
  By induction, the running time of the algorithm to process $A$ and $B$ 
  is less than $c. (n_A^C + n_B^C)$.
  Therefore, the overall time computation at node $N$ is less than $c. (n_A^C + n_B^C + (n_A^{C/2} * n_B^{C/2})) \le c . (n_A + n_B)^C = n^C$.

  We conclude that the algorithm takes $\mathcal{O}(n^C)$ computation time.
\end{proof}
\end{lem}

\subsection{Maximum Clique Problem}
We now tackle the maximum clique problem, namely:
\begin{defn}{Maximum Clique Problem.}\\
\emph{Instance:} A graph $G = (V,E)$.\\
\emph{Problem:} Find the size of a maximum complete subgraph of $G$.
\end{defn}

Let $G$ be a switch cograph, $N = (N_1, N_2)$ be a node of the tree and $A = (A_1,A_2),B = (B_1,B_2)$ be its two children
.



\begin{lem}{\textbf{Maximum Clique on switch cographs}. Let $G$ be a switch cograph, $N$ be a node of its involution modular binary 
    IMDT and $A = (A_1,A_2)$ and $B = (B_1,B_2)$ be its two children.
    \begin{description}      
    \item[If $N$ is a clique node], then the maximum clique of $G$ is the maximum clique among the maximum
      clique of $N_1$ , the maximum clique of $N_2$ , the maximum clique of $A$ and the maximum clique of $B$.
    \item[If $N$ is a bipartite node], then the maximum clique of G is the maximum clique among the maximum
      clique of $G_{[A_1 \cup B_2 ]}$ , the maximum clique of $G_{[A_2 \cup B_1 ]}$ , the maximum clique of $A$ and the maximum
      clique of $B$.
    \end{description}
  }
  \label{max_clique_switch cographs}
  \begin{proof}
    We first give the proof for the clique case.
    The maximum clique of $G_{[N]}$ cannot contain simultaneously elements of $A_1$ and $B_2$ nor simultaneously
    elements of $B_1$ and $C_2$ (since there is no edge between these elements). 
    Therefore, if the maximum clique contains elements of $A_i$ (resp $B_j$) 
    then it can contain elements of $B_i$ (resp $A_j$) and in this case no element of $A_{3-i}$ (resp $B_{3-j}$) or the converse 
    ($i \in \{1,2\}$).

    Let us now consider the bipartite case.
    The maximum clique of $G_{[N]}$ cannot contain simultaneously elements of $A_1$ and $B_1$ nor simultaneously
    elements of $A_2$ and $B_2$ (since there is no edge between these elements).
    Therefore, if the maximum clique contains elements of $A_i$ (resp $B_j$) 
    then it can contain elements of $B_j$ (resp $A_i$) and in this case no element of $A_{3-i}$ (resp $B_{3-j}$) or the converse 
    ($i \in \{1,2\}$).
  \end{proof}
\end{lem}

We now show that we are able to compute these values
for each node of the binary IMDT (assuming we already computed the corresponding values for its children).

\begin{thm}{The maximum clique problem can be solved with a linear time complexity when restricted to switch cographs.}
\begin{proof}
  Going from the leaves of the tree to its arbitrary-chosen root, for each node $N$ with $A=(A_1,A_2)$ and $B=(B_1,B_2)$ its
  two children. 
  We distinguish the two cases:
  \begin{description}
  \item[Clique Case:] If $N = (N_1,N_2)$ is a clique node such that $A_1$,$B_1$ and $A_2,B_2$ are completely joined.
    We first compute the size of the largest clique of $G_{[N_1]}$ (resp. $G_{[N_2]}$), which is, by lemma \ref{max_clique_switch cographs} 
    sum of the size of the largest clique of $G_{[A_1]}$ 
    (resp. $A_2$) and the size of the largest clique of $G_{[B_1]}$ (resp. $G_{[B_2]}$). 
    This operation can be done in $\mathcal{O}(1)$ by lemma \ref{crucial_lemma}.
    We can now compute the maximum clique of $G_{[N]}$ 
    which is the maximum among the maximum clique of $G_{[N_1]}$, the maximum clique of $G_{[N_2]}$,
    the maximum clique of $G_{[A]}$ and the maximum clique of $G_{[B]}$. 
    This operation can be done in $\mathcal{O}(1)$ provided we already computed the largest clique of its children 
    and by lemma \ref{crucial_lemma}.
    
  \item[Bipartite Case:] If $N = (N_1,N_2)$ is a bipartite node such that $A_1$,$B_2$ and $A_2$,$B_1$ are completely joined.
    We first compute the size of the largest clique on $G_{[N_1]}$ (resp. $G_{[N_2]}$), which is, by lemma \ref{max_clique_switch cographs}
    the maximum of the size of the largest clique of $G_{[A_1]}$ 
    (resp. $A_2$) and the size of the largest clique of $G_{[B_1]}$ (resp. $G_{[B_2]}$). 
    This operation can be done in $\mathcal{O}(1)$ by lemma \ref{crucial_lemma}.
    We can now compute the maximum clique of $G_{[N]}$ 
    which is the maximum among the maximum clique of $G_{[N_1]}$, the maximum clique of $G_{[N_2]}$,
    the maximum clique of $G_{[A]}$ and the maximum clique of $G_{[B]}$. 
    This operation can be done in $\mathcal{O}(1)$ provided we already computed the largest clique of its children
    and by lemma \ref{crucial_lemma}.

  \end{description}
  
  There are $\mathcal{O}(n)$ nodes on the tree-decomposition, the tree-decomposition can be computed
  in $\mathcal{O}(n+m)$, this leads to an $\mathcal{O}(n+m)$ algorithm to compute the maximum clique of a switch cograph.
\end{proof}

\end{thm}

\begin{cor}{The maximum independant set problem can be solved with a linear time complexity when restricted to switch cographs.}
\begin{proof}
The same reasoning applies when computing an independant set.
\end{proof}
\end{cor}

\begin{cor}{The chromatic number problem can be solved with a linear-time complexity when restricted to switch cographs.}
\begin{proof}
The class of switch cographs is included in the class of perfect graphs.
\end{proof}
\label{chrom_switch cographs}
\end{cor}

\begin{cor}{The Clique Cover and Independant Set Cover problems can be solved with a linear time complexity when restricted to switch cographs.}
\begin{proof}
The class of switch cographs is closed under complement and the chromatic number of a switch cograph can be computed in linear time.
\end{proof}

\end{cor}





\subsection{Vertex Cover Problem}
Let us now address the minimum vertex cover problem on switch cographs.
\begin{defn}{Minimum Vertex Cover Problem.}\\
\emph{Instance:} A graph $G = (V,E)$.\\
\emph{Problem:} Find a set of vertices $X$ such that $\forall \{x,y\} \in E$, $x \in X$ or $y \in X$ whose size is minimum.
\end{defn}

We first introduce the two following lemmas.

\begin{lem}{\textbf{Vertex Cover Problem on complete bipartite subgraphs}. Let $G=(A,B)$ be a complete bipartite 
    subgraph of a graph $H = (V,E)$. Let $S$ be a solution to the Vertex Cover Problem for $H$. 
    Then, either $A \subseteq S$ or $B \subseteq S$.}
\label{vcp_bipartite}

\begin{proof}
Suppose that neither $A$ or $B$ is included in $S$. Then there exists $a \in A$, $b \in B$ such that $a,b \notin S$.
But because $G=(A,B)$ is a complete bipartite subgraph, it means that the edge $(a,b) \in E$ is not covered by $S$,
a contradiction which concludes the proof.
\end{proof}
\end{lem}


\begin{lem}{\textbf{Vertex Cover Problem for Switch Cographs}. Let $G$ be a switch cograph, $N = (N_1,N_2)$ a node of its binary IMDT
and $A = (A_1,A_2)$ and $B = (B_1,B_2)$ be its two children.
The minimal Vertex Cover for $G_{[N]}$ is the minimal solution among:
\begin{description}
\item[If $N$ is a clique node]:
\begin{description}
\item[(1)] $S_1 = A_1 \cup A_2 \cup S_B$; 
\item[(2)] $S_2 = B_1 \cup B_2 \cup S_A$;
\item[(3)] $S_3 = A_1 \cup B_2 \cup S_{B_1} \cup S_{A_2}$;
\item[(4)] $S_4 = A_2 \cup B_1 \cup S_{B_2} \cup S_{A_1}$;
\end{description}

\item[If $N$ is a bipartite node]:
\begin{description}
\item[(1)] $S_5 = A_1 \cup A_2 \cup S_B$;
\item[(2)] $S_6 = B_1 \cup B_2 \cup S_A$;
\item[(3)] $S_7 = A_1 \cup B_1 \cup S_{B_2} \cup S_{A_2}$;
\item[(4)] $S_8 = A_2 \cup B_2 \cup S_{B_1} \cup S_{A_1}$;
\end{description}

where $S_A$, $S_B$, $S_{A_1}$, $S_{A_2}$, $S_{B_1}$ and $S_{B_2}$ are respectively vertex cover solutions to $G_{[A]}$, $G_{[B]}$, $G_{[A_1]}$, $G_{[A_2]}$, $G_{[B_1]}$ and $G_{[B_2]}$.

\end{description}
}
\label{vcp_switch cograph}
\begin{proof}
First, notice that cases (1) and (2) are symmetric (and (3) and (4) as well) regardless of whether $N$ is a clique or a bipartite node.

\begin{description}
\item[If $N$ is a clique node]:
By lemma \ref{vcp_bipartite}, either $A_1$ or $B_1$ (resp $A_2$ or $B_2$) is included in $S$.
We consider the two possible cases (up to symmetry), $A_1 \cup A_2 \in S$ or $A_1 \cup B_2 \in S$.

In the first case, it is guaranted that all the edges between elements of $A$ and all the edges between $A$ and $B$ are covered, 
therefore we just need to cover the edges of $B$ and we use the solution for $G_{[B]}$ to cover them. This proves 
the cases (1) (and (2) likewise).

In the second case, it is guaranted that all the edges between $A_1$ (resp $B_2$) and $B_1 \cup A_2$ are covered. We just need
to cover the edges between elements of $B_1$ and the edges between elements of $A_2$. This proves cases (3) (and (4) likewise).

Since these cases are the only possible cases, this completes the clique node case proof.

\item[If $N$ is a bipartite node]:
By lemma \ref{vcp_bipartite}, either $A_1$ or $B_1$ (resp $A_2$ or $B_2$) is included in $S$.
We consider the two possible cases (up to symmetry), $A_1 \cup A_2 \in S$ or $A_1 \cup B_2 \in S$.

In the first case, it is guaranted that all the edges between elements of $A$ and all the edges between $A$ and $B$ 
are covered, 
therefore we just need to cover the edges of $B$ and we use the solution for $G_{[B]}$ to cover them. This proves 
the cases (1) (and (2) likewise).

In the second case, it is guaranted that all the edges between $A_1$ (resp. $B_1$) and $B_2 \cup A_2$ (resp. $A_2 \cup B_2$)
are covered. We just need to cover the edges between elements of $B_2$ and the edges between 
elements of $A_2$. this proves the cases (3) and ((4) likewise).

Since these cases are the only possible cases, this completes the bipartite node case proof.

\end{description}
\end{proof}
\end{lem}

\begin{thm}{The Vertex Cover Problem can be solved with a linear time complexity when restricted to switch cographs.}

\begin{proof}
We propose a bottom-up algorithm, from the leaves of the tree to its root.
We claim that we are able to compute for each node $N=(N_1,N_2)$ of the tree, an optimal solution to Vertex Cover
for $N$, $N_1$, and $N_2$ in constant time provided the solutions to the vertex cover problem on the children of $N$.

This is obviously true for the leaves since there is only one node.
Assume that this holds for any node at a distance of at most $k$ from the root and we show that this holds
for a node at a distance $k-1$.

Let $N = (N_1,N_2)$ be such a node and $A=(A_1,A_2)$ and $B = (B_1,B_2)$ be its two children.

\begin{description}
\item[Clique Node:] If $N$ is a clique node. We first show that we are able to compute $S_{N_1}$ (resp $S_{N_2}$) a solution 
to vertex cover to the subgraph $G_{[N_1]}$ (resp $G_{[N_2]}$). By lemma \ref{vcp_bipartite}, 
either $A_1$ or $B_1$ is included in $S_{N_1}$.

If all the elements of $A_1$ are included then we just need to cover the edges of $G_{[B_1]}$ so we add $A_1$ to the solution 
to vertex cover for $G_{[B_1]}$. By induction hypothesis and lemma \ref{crucial_lemma} we already computed this solution.
If all the elements of $B_1$ are included we just need to cover the edges of $G_{[A_1]}$ so we add $B_1$ to the solution to vertex cover
for $G_{[A_1]}$. By induction hypothesis and lemma \ref{crucial_lemma} we also computed this solution.

These cases are the two possible cases, so the solution to vertex cover for $G_{[N_1]}$ is the smallest
among these two cases. We can do the same operation to compute an optimal solution for $G_{[N_2]}$.

\item[Bipartite Node:] If $N$ is a bipartite node. We first show that we are able to compute $S_{N_1}$ (resp $S_{N_2}$) a solution 
to vertex cover to the subgraph $N_1$ (resp $N_2$). Since there is no edge between $G_{[A_i]}$ and $G_{[B_i]}$, 
the  solution for
$G_{[N_i]}$ is $S_{A_i} \cup S_{B_i}$ where $S_{A_i}$ and $S_{B_i}$ are the solution for $G_{[A_i]}$ and $G_{[B_i]}$.
By induction hypothesis and lemma \ref{crucial_lemma}, these solutions are already computed.
\end{description}

We now show that we are able to compute a solution for $G_{[N]}$. By lemma \ref{vcp_switch cograph} there are only 4 different
cases and by induction hypothesis and lemma \ref{crucial_lemma} we have already computed the solutions for 
$G_{[A_1]},G_{[A_2]}, G_{[B_1]},G_{[B_2]},G_{[A]}$ and $G_{[B]}$. Thus we can choose the smallest among the four cases we described 
above to be the solution for $G_{[N]}$.

We now prove the complexity and correctness of the algorithm.
\begin{description}
\item[Correctness]: By lemma \ref{vcp_switch cograph}, \ref{vcp_bipartite}, \ref{crucial_lemma}.
\item[Complexity]: At each node, we compute the minimum over 4 values. The tree has $\mathcal{O}(n)$ nodes, the tree-decomposition is
  computed in $\mathcal{O}(n+m)$ the complexity of the algorithm is therefore $\mathcal{O}(n + m)$.
\end{description}

\end{proof}
\end{thm}

\subsection{Maximum Cut Problem}
Whereas the clique-width of switch cographs is bounded (see section \ref{clique-width section}), the complexity of the maximum cut problem remained open.
This problem is not expressible in monadic second order logic (\cite{algo_lower_bound}) and
therefore is not caught by the famous theorem of Courcelle et al. \cite{courcelle_msol}.
This shows that our decomposition tool provides us with even more algorithmic properties than the clique-width decomposition for the class of switch cographs.

\begin{defn}{Maximum Cut Problem.}\\
\emph{Instance:} A graph $G = (V,E)$.\\
\emph{Problem:} Find two sets of vertices $A$ and $B$ such that $A \cap B = \emptyset$ and the set $\{\{x,y\}$ $|$ $\{x,y\} \in E$, $x \in A$ and $y \in B \}$ 
has maximum size.
\end{defn}

In order to compute the Maximum Cut of a switch cograph we will use a dynamic programming approach.
We first prove two lemmas of particular importance.

\begin{lem}{\textbf{Maximum Cut on Switch Cographs}.
    Let $G=(V,E)$ be a switch cograph and $N = (N_1,N_2)$ a node of its IMDT and $A=(A_1,A_2)$ 
    and $B=(B_1,B_2)$ its two childen. Let us define $C_N$ be a $|V| \times |V|$ array such that $C_N[i,j]$ equals 
    the size of a maximum cut $(X,Y)$ such that $|X \cap N_1| = i$ and $|X \cap N_2| = j$.    
    Then\\
\begin{description}
\item[If $N$ is a clique node]:\\
    $C_N[i,j] = \max\limits_{\scriptsize \begin{array}{c} k+l=i \\ q+r=j \end{array}} C_A[l,q] + C_B[k,r] + l(|B_1|-k) + k(|A_1|-l) + q(|B_2|-r) + r(|A_2|-q)$
\item[If $N$ is a bipartite node]:\\
    $C_N[i,j] = \max\limits_{\scriptsize \begin{array}{c} k+l=i \\ q+r=j \end{array}} C_A[l,q] + C_B[k,r] + 
    q(|B_1|-k) + k(|A_2|-q) + l(|B_2|-r) + r(|A_1|-l)$

\end{description}
}
\label{mcp_switch cograph}
\begin{proof}
We distinguish the two sorts of nodes. 
\begin{description}
\item[If $N$ is a clique node]
Let $S=(X,Y)$ be a maximum cut of value $v$ such that $|X \cap N_1| = i$, $|X \cap N_2| = j$, $|X \cap A_1| = l$, $|X \cap A_2| = q$,
$|X \cap B_1| = k$ and $|X \cap B_2| = r$.

Assume towards contradiction that $v > C_N[i,j] = C_A[l,q] + C_B[k,r] + l(|B_1|-k) + k(|A_1|-l) + q(|B_2|-r) + r(|A_2|-q)$.

Since there are $l$ elements from $A_1$ in $X$ and $k$ elements from $B_1$ in $X$, there are $l(|B_1|-k) + k(|A_1|-l)$ edges
crossing the cut in $G_1$. The same reasoning applies to $G_2$.

Now, let $v_A$ (resp. $v_B$) be the value of the cut $S$ restricted to the edges of $G_{[A]}$ (resp. $G_{[B]}$). Since $v$ is strictly greater than $C_N[i,j]$ it means that
$v_A + v_B > C_A[l,q] + C_B[k,r]$.
By induction hypothesis and lemma \ref{crucial_lemma},
we have both $C_A[l,q] \ge v_A$ and $C_B[k,r] \ge v_B$, a contradiction which concludes the proof.

\item[If $N$ is a bipartite node]
Let $S=(X,Y)$ be a maximum cut of value $v$ such that $|X \cap N_1| = i$, $|X \cap N_2| = j$, 
$|X \cap A_1| = l$, $|X \cap A_2| = q$, $|X \cap B_1| = k$ and $|X \cap B_2| = r$.

Assume towards contradiction that $v > C_N[i,j] = C_A[l,q] + C_B[k,r] + q(|B_1|-k) + k(|A_2|-q) + l(|B_2|-r) + r(|A_1|-l)$.

Since there are $l$ elements from $A_1$ in $X$ and $r$ elements from $B_2$ in $X$, there are exactly
$l(|B_2|-r) + r(|A_1|-l)$ edges crossing the cut from $G_{[A_1]}$ to $G_{[B]}$. We apply the same reasoning to $A_2$.

Now, let $v_A$ (resp. $v_B$) be the value of the cut $S$ restricted to the edges of $G_{[A]}$ (resp. $G_{[B]}$). 
Since $v$ is strictly greater than $C_N[i,j]$ it means that $v_A + v_B > C_A[l,q] + C_B[k,r]$.
By induction hypothesis and lemma \ref{crucial_lemma},
we have both $C_A[l,q] \ge v_A$ and $C_B[k,r] \ge v_B$, a contradiction which concludes the proof.

\end{description}

We now show that if $C_N[i,j] = v$ there exists a cut $S$ such that $|X \cap N_1| = i$ and $|X \cap N_2| = j$ of value $v$.
There exist $l,q,k,r \in \{0,...,n\}$ such that $v = C_A[l,q] + C_B[k,r] + l(|B_1|-k) + k(|A_1|-l) + q(|B_2|-r) + r(|A_2|-q)$.
By induction hypothesis, there exist cuts of $A$ and $B$ such that $|X \cap A_1| = l$, $|X \cap A_2| = q$, $|X \cap B_1| = k$ and $|X \cap B_2| = r$ of values
$C_A[l,q]$ and $C_B[k,r]$.
By combining these two cuts, one obtain a new cut of value $v$ in both cases.

\end{proof}

\end{lem}






\begin{thm}{The Maximum Cut Problem can be solved in  $\mathcal{O}(n^4)$ time when restricted to switch cographs.}
\begin{proof}
\begin{description}
\item[Correctness]: By lemmas \ref{mcp_switch cograph} and \ref{crucial_lemma}.
\item[Complexity]: 
  

  The computing time at a single node of the tree is $(n_A^2 * n_B^2)$ where $n_A$ and $n_B$ are respectively the number of nodes
  in the subtrees rooted on $A$ and $B$.
  We conclude by lemma \ref{lem_complexity} that the algorithm takes an $\mathcal{O}(n^4)$ computation time.
\end{description}
\end{proof}
\end{thm}








\subsection{Vertex Separator Problem}

In this section we show that the vertex separator problem is polynomially tractable when restricted to switch cographs.

\begin{defn}{Vertex Separator Problem.}\\
\emph{Instance}: $G=(V,E)$.\\
\emph{Problem}: Find sets $X_1,X_2 \subseteq V$, such that\\
(1) $X_1 \cup X_2 = V$.\\
(2) For each $\{u,v\} \in E$ there is an $i \in \{1,2\}$ s.t $\{u,v\} \subseteq X_i$.\\
and minimize $\max(|X_1|,|X_2|)$.

\end{defn}

\begin{thm}{\cite{vsp_npc}. The Vertex Separator problem is NP-Complete.}
\end{thm}

We first show that the Vertex Separator problem is polynomial when restricted to the class of cographs and
we prove the following lemma.

\begin{lem}{\textbf{Bipartite subgraph contention}.
    Let $G=(V,E)$ be a graph and $(A,B)$ be a complete bipartite subgraph of $G$.
    Let $X_1,X_2$ be an optimal solution for the vertex separator problem on $G$.
    If $A \not\subseteq X_1$ and $A \not\subseteq X_2$ then $B \subseteq X_1$ and $B \subseteq X_2$.
}
\label{vsp_bipartite}
\begin{proof}
Assume that $A \not\subseteq X_1$ and $A \not\subseteq X_2$, then there exists $a_1,a_2 \in A$ such that
$a_1 \in X_1$, $a_1 \notin X_2$ and $a_2 \in X_2$, $a_2 \notin X_1$. These two vertices are both adjacent to all the 
vertices of $B$, therefore, in order to fullfil (2) $X_1$ and $X_2$ have to contain $B$.\\
\end{proof}

\end{lem}

In the following, we refer to $X_1$ and $X_2$ as ``bags''.

\begin{thm}{The Vertex Separator Problem can be solved in $\mathcal{O}(n^4)$ time when restricted to cographs.}

\label{thm_vsp_cograph}

\begin{proof}
Let $G=(V,E)$ a cograph and $T$ a binary modular decomposition tree of $G$.
We compute a solution to the vertex separator problem in a bottom-up fashion, i.e going from the leaves to the root of
the tree. We compute a solution for a node when its two children have been computed.

We use dynamic programming in the following sense: for each node $N$ of the tree and its children $C_1$ and $C_2$,
we define an $|V| \times |V|$ boolean array $Z_N$ such
that $Z_N[i,j]$ iff there exists a solution to the vertex separator restricted to $G_{[N]}$ with $|X_1| = i$ and $|X_2| = j$. 
Consider a step at a node $N$ and assume that we already filled the two arrays of its two children, $Z_{C_1}$ and $Z_{C_2}$.
We distinguish the two following cases:

\begin{enumerate}
\item First, if the node $N$ is a series node, namely there are all the edges between the vertices of $C_1$ and the vertices of $C_2$.
  By lemma \ref{vsp_bipartite}, if $C_1$ and $C_2$ are not empty we have either $C_1$ on the two bags or $C_2$ on the two bags.
  In the first case, the minimal vertex separator consists of taking the minimal solution of $C_2$ and adding
  $C_1$ on both bags. In the second case, take the minimal solution of $C_1$ and add $C_2$ on both bags.
  We now fill the array in the following way: $\forall i,j \in \{0,...,n\}$, $Z[i,j] \iff Z_{C_1}[i - |C_2|, j - |C_2|]$ or
  $Z_{C_2}[i - |C_1|, j - |C_1|]$.

\item Let us now suppose $N$ is a parallel node, namely there is no edge between the vertices of $C_1$ and the vertices of $C_2$.
  We can fill the array in the following way: 
  $\forall i,j \in \{0,...,|V|\}$, $Z[i,j] \iff \exists k,l,f,h \in \{1,...,n\}$, $k+l = i$ and $f+h=j$, such that 
  $Z_{C_1}[k,f]$ and $Z_{C_2}[l,h]$ or $Z_{C_1}[k,f]$ and $Z_{C_2}[h,l]$ or $Z_{C_1}[f,k]$ and $Z_{C_2}[l,h]$
  or $Z_{C_1}[f,k]$ and $Z_{C_2}[h,l]$. We mean here that for any solutions of $C_1$ and $C_2$, say $(X_{C_1}^1,X_{C_1}^2)$
  and $(X_{C_2}^1,X_{C_2}^2)$, $(X_{C_1}^i \cup X_{C_2}^j, X_{C_1}^k \cup X_{C_2}^l)$ are solutions for $N$ for any $i,j,k,l \in \{1,2\}$, 
  $i \neq k$ and $j \neq l$. 
\end{enumerate}
\begin{description}
\item[Correctness:] 
  We show that the algorithm we gave above is correct. In the parallel case, let $S = (X,Y)$ be a solution, then 
  $S$ restricted to $A$ and $S$ restricted to $B$ are solutions to $A$ and $B$.
  In the series case, let $S=(X,Y)$ be an optimal solution, then by lemma \ref{vsp_bipartite}, this solution contains either $A$ or
  $B$. In the first case, the solution restricted to $B$ is a solution for $B$ and the same reasoning applies to $A$. 
\item[Complexity:] 
  The computing time at a single node of the tree is $(n_A^2 * n_B^2)$ where $n_A$ and $n_B$ are respectively the number of nodes
  in the subtrees rooted on $A$ and $B$.
  We conclude by lemma \ref{lem_complexity} that the algorithm takes an $\mathcal{O}(n^4)$ computation time.

\end{description}
\end{proof}
\end{thm}

We now show that the Vertex Separator problem is polynomial when restricted to the class of switch cographs.

\begin{thm}{The Vertex Separator problem can be solved in $\mathcal{O}(n^8)$ time when restricted to switch cographs.}
\begin{proof}

Let $G = (V,E)$ be a switch cograph and $T$ the binary IMDT of $G$.
For each node $N = (N_1,N_2)$ of the tree and its children $A=(A_1,A_2)$ and $B=(B_1,B_2)$, we define an $n \times n \times n \times n$ 
boolean array $Z_N$, such that $Z_N[X_{N_1}, X_{N_2}, Y_{N_1}, Y_{N_2}]$ if and only if there exists a solution $(X,Y)$ 
to the vertex separator problem on $G_{[N]}$ such that
$|N_1 \cap X| = X_{N_1}$, $|N_2 \cap X| = X_{N_2}$, $|N_1 \cap Y| = Y_{N_1}$, $|N_2 \cap Y| = Y_{N_2}$.
If we are able to fill this array, we can find the optimal solution among the possible values of $X_{N_1},X_{N_2},Y_{N_1},Y_{N_2}$,
i.e the solution which minimizes $\max(X_{N_1} + X_{N_2}, Y_{N_1} + Y_{N_2})$.

Now we give the following algorithm to fill this array:\\

\begin{description}
\item[Clique Node:]
  $Z_G[X_{N_1}, X_{N_2}, Y_{N_1}, Y_{N_2}]$ $\iff$ $\exists X_{A_1},X_{A_2},X_{B_1},X_{B_2},Y_{A_1},Y_{A_2},Y_{B_1},Y_{B_2}$, such that
  $X_{A_1} + X_{B_1} = X_{N_1}$, $X_{A_2} + X_{B_2} = X_{N_2}$, $Y_{A_1} + Y_{B_1} = Y_{N_1}$, $Y_{A_2} + Y_{B_2} = Y_{N_2}$, 
  \begin{description}
  \item[(1)] $Z_A[X_{A_1},X_{A_2},Y_{A_1},Y_{A_2}]$ and $Z_B[X_{B_1},X_{B_2},Y_{B_1},Y_{B_2}]$ and 
  \item[(2)] $\forall,i,j \in \{1,2\}$, $i \neq j$, if $X_{A_i} < |A_i|$ then $X_{B_i} = |B_i|$ and if $X_{B_j} < |B_j|$ then $X_{A_j} = |A_j|$ and 
    if $Y_{A_i} < |A_i|$ then $Y_{B_i} = |B_i|$ and if $Y_{B_j} < |B_j|$ then $Y_{A_j} = |A_j|$. 
  \end{description}

\item[Bipartite Node:]
  $Z_G[X_{N_1}, X_{N_2}, Y_{N_1}, Y_{N_2}]$ $\iff$ $\exists X_{A_1},X_{A_2},X_{B_1},X_{B_2},Y_{A_1},Y_{A_2},Y_{B_1},Y_{B_2}$, such that
  $X_{A_1} + X_{B_1} = X_{N_1}$, $X_{A_2} + X_{B_2} = X_{N_2}$, $Y_{A_1} + Y_{B_1} = Y_{N_1}$, $Y_{A_2} + Y_{B_2} = Y_{N_2}$, 
  \begin{description}
  \item[(1)] $Z_A[X_{A_1},X_{A_2},Y_{A_1},Y_{A_2}]$ and $Z_B[X_{B_1},X_{B_2},Y_{B_1},Y_{B_2}]$ and 
  \item[(2)] $\forall,i,j \in \{1,2\}$, $i \neq j$, if $X_{A_i} < |A_i|$ then $X_{B_j} = |B_j|$ and if $X_{B_i} < |B_i|$ then $X_{A_j} = |A_j|$ and 
    if $Y_{A_j} < |A_j|$ then $Y_{B_i} = |B_i|$ and if $Y_{B_i} < |B_i|$ then $Y_{A_j} = |A_j|$. 
  \end{description}
\end{description}

\begin{description}

\item[Correctness:]
  We give the proof for a clique node, the same reasoning applies for a bipartite node.
  We first prove that for any clique node $N=(N_1,N_2)$, $\forall X_{N_1}, X_{N_2}, Y_{N_1}$, $Y_{N_2} \in \{1,...,|V|\}$,\\ 
  $Z_G[X_{N_1}, X_{N_2}, Y_{N_1}, Y_{N_2}]$ $\iff$ there exists a solution to the vertex separator problem for $G$ with 
  $|N_1 \cap X| = X_{N_1}, |N_2 \cap X| = X_{N_2}, |N_1 \cap Y| = Y_{N_1}, |N_2 \cap Y| = Y_{N_2}$. 
  This is obviously true for the leaves and we assume that it holds for each node at a distance of at most $k$ from the root.
  We now show that it holds for any node at a distance $k-1$. We assume we already filled the array for its two children
  $A=(A_1,A_2)$ and $B=(B_1,B_2)$.

  By induction hypothesis, we assume that $\forall X_{A_1}$, $X_{A_2}$, $Y_{A_1}$, $Y_{A_2}$ $\in$ $\{1,...,|V|\}$,\\ 
  $Z_A[X_{A_1}$,$ X_{A_2}$,$ Y_{A_1}$,$ Y_{A_2}]$ $\iff$ there exists a solution $(X',Y')$ to the vertex separator problem on $A$ such that 
  $|A_1 \cap X'| = X_{A_1}, |A_2 \cap X'| = X_{A_2}, |A_1 \cap Y'| = Y_{A_1}, |A_2 \cap Y'| = Y_{A_2}$
  and the same applies to $B$.
  
  \begin{description}
    \item[(1)] $Z_A[X_{A_1},X_{A_2},Y_{A_1},Y_{A_2}]$, $Z_B[X_{B_1},X_{B_2},Y_{B_1},Y_{B_2}]$ and 
    \item[(2)] $forall,i,j \in \{1,2\}$, if $X_{A_i} < |A_i|$ then $X_{B_i} = |B_i|$ and if $X_{B_j} < |B_j|$ then $X_{A_j} = |A_j|$ and 
      if $Y_{A_i} < |A_i|$ then $Y_{B_i} = |B_i|$ and if $Y_{B_j} < |B_j|$ then $Y_{A_j} = |A_j|$. 

  \end{description}
  By lemma \ref{vsp_bipartite}, if the solution fullfils the requirement (2) then all the edges between $B_i$ and $A_i$ are covered.
  By induction hypothesis, if the solution fullfils the requirement (1) then all the edges between $A_1$ and $A_2$ (resp $B_1$ and $B_2$) 
  are covered.
  We can conclude that $\forall X_{N_1}, X_{N_2}, Y_{N_1}, Y_{N_2} \in \{1,...,|V|\}, 
  Z_G[X_{N_1}, X_{N_2}, Y_{N_1}, Y_{N_2}] \implies$ there exists a solution to the vertex separator problem for $G$ with 
  $|N_1 \cap X| = X_{N_1}, |N_2 \cap X| = X_{N_2}, |N_1 \cap Y| = Y_{N_1}, |N_2 \cap Y| = Y_{N_2}$. 
  
  We now show the converse, let $(X,Y)$ be a solution to the vertex separator problem for $G$ such that
  $|N_1 \cap X| = X_{N_1}, |N_2 \cap X| = X_{N_2}, |N_1 \cap Y| = Y_{N_1}, |N_2 \cap Y| = Y_{N_2}$ and
  there exists $X_{A_1},X_{A_2},X_{B_1},X_{B_2},Y_{A_1},Y_{A_2},Y_{B_1},Y_{B_2}$, such that 
  $X_{A_1} + X_{B_1} = X_{N_1}$, $X_{A_2} + X_{B_2} = X_{N_2}$, $Y_{A_1} + Y_{B_1} = Y_{N_1}$, $Y_{A_2} + Y_{B_2} = Y_{N_2}$, and 
  assume towards contradiction that the solution does not satisfy either (1) or (2) (for the clique node).

  By lemma \ref{vsp_bipartite}, the solution has to fullfil (2).

  If the solution does not satisfy (1), then w.l.o.g $Z_A[X_{A_1},X_{A_2},Y_{A_1},Y_{A_2}]$ is false.
  Therefore, by induction hypothesis, $(X \cap A, Y \cap A)$ is not a solution to the vertex separator
  problem for $A$, which means that there is an edge of $A$ which is not covered by the solution $(X,Y)$, a contradiction.




  It remains now to go through the array of the root in order to find the optimal solution.
  We can conclude that the algorithm computes the minimal vertex separator solution for $G$.

\item[Complexity:] 
  The computing time at a single node of the tree is $(n_A^4 * n_B^4)$ where $n_A$ and $n_B$ are respectively the number of nodes
  in the subtrees rooted on $A$ and $B$.
  We conclude by lemma \ref{lem_complexity} that the algorithm takes an $\mathcal{O}(n^8)$ computation time.
\end{description}

\end{proof}
\end{thm}

\subsection{Clique-Width}
\label{clique-width section}
Let us now consider the clique-width problem, namely:

\begin{defn}{Clique-Width number.\\
The Clique-Width number of a graph is the minimum number of different labels that is needed to construct the graph using the following operations:
\begin{enumerate}
\item Creation of a vertex with label $i$,
\item Disjoint union of two graphs,
\item Relabelling the nodes labeled $i$ with label $j$,
\item Connecting all vertices with label $i$ to all vertices with label $j$.
\end{enumerate}}
\end{defn}

\begin{figure}
\begin{center}
\includegraphics[scale=0.5]{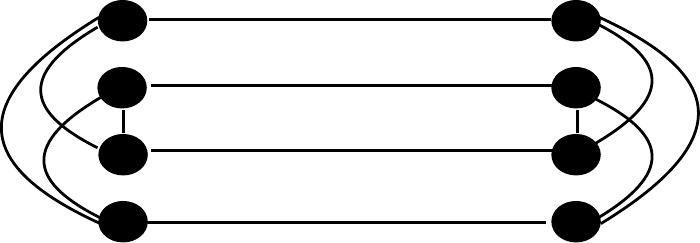}
\caption{
  A switch cograph with Clique-Width 4.
}
\label{cw4}

\end{center}
\end{figure}

\begin{thm}{The class of switch cographs in strictly included in the class of 
    graphs with Clique-Width at most 4.
}
\label{cw4_thm}
\begin{proof}
We first show that any switch cographs has Clique-Width at most 4.
Let $G=(V,E)$ be a switch cograph. Build the binary IMDT of $G$.
We proceed bottom-up to build the graph with 4 labels.

We show that for a node $N=(N_1,N_2)$, it is possible to construct the graph $G_{[N]}$
with at most 4 labels and such that $N_1$ receive at most 2 labels and $N_2$ as well and
$N_1$ and $N_2$ do not share any label.
Clearly it is always possible for the leaves, we assume this is possible for each
node at a distance of at most $k+1$ from the root and we show that it holds for the nodes at 
a distance $k$. We assume we already built $A$ and $B$.

Let $N=(N_1,N_2)$ be such a node and $A=(A_1,A_2)$ and $B=(B_1,B_2)$ be its two children.
By induction hypothesis $A_1$ has at most 2 labels, w.l.o.g \emph{label1} and \emph{label2},
and $A_2$ has no vertex with these labels; $A_2$ has only vertices with \emph{label3} and \emph{label4},
We first relabel the vertices labeled \emph{label2} with \emph{label1} such that $A_1$ has only
vertices labeled \emph{1}.   

Now, we process $B_1$ by relabelling the vertices of $B_1$ in such way that they all receive \emph{label2}.

We now relabel the vertices of $A_2$ in such a way that they all receive \emph{label3}
and the vertices of $B_2$ in such way that they all receive \emph{label4}.

Remark that $A_1$,$A_2$,$B_1$,$B_2$ received labels that are pairwise different.

We now make the disjoint union of $A$ and $B$.
We use then the (4) rule to connect
the vertices of $A_1$ to the vertices of $B_1$ (resp. $A_2$ to $B_2$) if the node is a clique node and
the vertices of $A_1$ to the vertices of $B_2$ (resp. $A_2$ to $B_1$) if the node is a bipartite node.

We created $G_{[N]}$ and we ensured that $N_1$ and $N_2$ received different colors 
and each at most two colors.\\

Figure \ref{cw4} shows that the bound is tight, there exists a switch cograph with 
Clique-Width 4.\\

The bull graph is a forbidden induced subgraph with Clique-Width 3 and so 
the class of switch cographs is strictly included in the class of graphs with Clique-Width
at most 4.
\end{proof}

\end{thm}

\begin{cor}{For any switch cograph, one can compute a clique-width expression of clique-width at most 4 in linear time.
  }

\begin{proof}
  As described in the proof of theorem \ref{cw4_thm}, the binary modular involution tree of any switch cograph leads to a clique-width expression
  of clique-width at most 4.
  This tree can be computed in $\mathcal{O}(n)$ time given the binary IMDT which can be computed in $\mathcal{O}(n+m)$.
\end{proof}

\end{cor}



\section{Concluding remarks and open problems}
We gave section \ref{involution_modules} a generalization of modular decomposition, the first which
has strong properties in more general contexts than graphs.
We showed that the family of involution modules has a unique linear-sized tree representation 
for any 2-structure. 
We derived an $\mathcal{O}(n^2)$ algorithm which computes the tree representation of the family of
involution modules of any 2-structure.

Then, we used our decomposition tool to demonstrate both algorithmic and structural properties on graphs.
We summarize our algorithmic results in table \ref{table}.
The table presents the classical graph problems we addressed with their previous best complexity and the complexity we obtain
thanks to our decomposition tool.
We then gave an algorithm to compute a Clique-Width expression of a Switch Cograph in linear time.
Thanks to the celebrated Courcelle's theorem \cite{courcelle_msol} this led all the problems expressible in MSOL$_1$ to be solved
in linear time.
Nevertheless the theorem induces a huge constant factor in the big-O notation that makes the algorithms impractical.
Based on our new framework, we gave easily implementable and optimal algorithms for these problems and we solved two other
problems that are not expressible in MSOL$_1$.
Besides, we also gave an $\mathcal{O}(n^4)$ algorithm for the vertex separator problem for cographs which was still open.
Eventually, we showed that the class of Switch Cographs has a clique-width bounded by 4. 

\begin{table}
\caption{A table of well-known graph problems and their current complexity for the class of switch cographs according to \cite{graphclasses}
and the complexity that result from our contribution.} 
\label{table}
\begin{center}
\begin{tabular}{|l|c|r|}
  \hline
  Problem & Current Best Result & Our Contribution\\
  \hline
  Maximum Clique            & Polynomial & $\mathcal{O}(n+m)$\\
  Maximum Independant Set   & Polynomial & $\mathcal{O}(n+m)$\\
  Minimum Clique Cover      & Polynomial & $\mathcal{O}(n+m)$\\
  Colourability             & Polynomial & $\mathcal{O}(n+m)$\\ 
  Recognition               & Polynomial & $\mathcal{O}(n+m)$\\
  Minimum Vertex Cover      & Polynomial & $\mathcal{O}(n+m)$\\
  Maximum Cut               & Unknown    & $\mathcal{O}(n^4)$\\
  Vertex Separator          & Unknown    & Polynomial\\
  Clique-Width              & 16         & 4\\ 
  Clique-Width Expression   & Polynomial & $\mathcal{O}(n+m)$\\
  \hline

\end{tabular}

\end{center}
\end{table}

We now present open questions.
Does our decomposition tool provides as many algorithmic properties as the modular decomposition tree?
Namely, most of the classical graph problems which are NP-complete in the general case have polynomial algorithms 
for the class of cographs thanks to the modular decomposition, does this hold for switch cographs?
For example, the complexity of the path cover problem is still open for the class of switch cographs.
Whereas the particular case of the hamiltonian path problem is caught by the Courcelle's theorem, the more general version 
of the problem, the path cover problem is not expressible in monadic second order logic \cite{cw6_pathcover}.
\cite{cograph_pathcover} presented a polynomial-time algorithm for the path cover
problem when restricted to cographs (the class of clique-width-2 graphs). Besides, \cite{cw6_pathcover} showed that 
this problem is NP-complete when restricted to the class of graphs with 
Clique-Width at most 6. The class of switch cographs is between these two classes.

Also, the tree-width of the switch cographs is not bounded and it is of particular interest to determine whether
computing the treewidth of a switch cographs can be computed in polynomial time or not. 

\begin{conj}{
    The problem of computing the tree-width of a switch cograph is NP-Complete.
  }
\end{conj}

\bibliographystyle{plain}
\bibliography{paper.bib}
\end{document}